\documentclass[sigconf, nonacm]{acmart}

\newcommand\vldbavailabilityurl{}

\usepackage{amsmath,amssymb,amsfonts,mathtools}
\usepackage{booktabs}
\usepackage{makecell}

\usepackage{tikz}
\usetikzlibrary{arrows.meta,positioning,calc,shapes.geometric}
\usepackage{graphicx}
\usepackage{balance}  
\usepackage{lipsum}
\usepackage{amsmath}
\usepackage{physics}
\usepackage{subfigure}
\usepackage{tabularx}
\usepackage{color}
\usepackage{colortbl}
\usepackage{float}
\usepackage{listings}
\usepackage{multirow}
\usepackage[normalem]{ulem}
 \usepackage{longtable}
\usepackage{enumitem}
\usepackage{multirow}
\usepackage{booktabs}
\usepackage{capt-of}
\usepackage{pifont}
\usepackage{ulem}
\usepackage{soul}
\usepackage{cancel}
\usepackage{bm}
\usepackage{url}
\usepackage{caption}
\usepackage[T1]{fontenc} 
\usepackage{hyperref}    
\usepackage{cleveref}
\crefformat{section}{\S\,#2#1#3} 
\crefformat{subsection}{\S\,#2#1#3}

\def\BibTeX{{\rm B\kern-.05em{\sc i\kern-.025em b}\kern-.08em
    T\kern-.1667em\lower.7ex\hbox{E}\kern-.125emX}}

 \graphicspath{{./Graph/}, {./Fig/}, {./Legend/}}

\long\def\comment#1{}

\newcounter{example}[section]
\renewcommand{\theexample}{\nthesection.\arabic{example}}

\newcounter{definition}[section]
\renewcommand{\thedefinition}{\nthesection.\arabic{definition}}

\newcounter{theorem}[section]
\renewcommand{\thetheorem}{\nthesection.\arabic{theorem}}
\newenvironment{theorem}{\begin{em}
        \refstepcounter{theorem}
        {\vspace{1ex} \noindent\bf  Theorem  \thetheorem:}}{
        \end{em}\vspace{1ex}} 

\newcounter{lemma}[section]
\renewcommand{\thelemma}{\nthesection.\arabic{lemma}}
\newenvironment{lemma}{\begin{em}
        \refstepcounter{lemma}
        {\vspace{1ex}\noindent\bf Lemma \thelemma:}}{
        \end{em}\vspace{1ex}} 

\newcounter{corollary}[section]
\renewcommand{\thecorollary}{\nthesection.\arabic{corollary}}

\newcounter{proposition}[section]
\renewcommand{\theproposition}{\nthesection.\arabic{proposition}}

\newcounter{remark}[section]
\renewcommand{\theremark}{\nthesection.\arabic{remark}}

\newcommand{\proofsketch}{\noindent{\bf Proof Sketch: }}

\newcommand{\nthesection}{\arabic{section}}

\newcommand{\eop}{\hspace*{\fill}\mbox{$\Box$}\vspace*{1ex}}

\newcommand{\stitle}[1]{\vspace{1mm} \noindent{\bf #1}}

\newcommand{\green}[1]{\textcolor{green}{}}

\newcommand{\kw}[1]{{\ensuremath {\mathsf{#1}}}\xspace}
\newcommand{\kwnospace}[1]{{\ensuremath {\mathsf{#1}}}}

\newcommand\bigutimes{\mathop{\ooalign{$\bigcup$\cr%
   \hfil\raise0.36ex\hbox{$\scriptscriptstyle\boldsymbol{\times}$}\hfil\cr}}}
\newcommand\bigumius{\mathop{\ooalign{$\bigcup$\cr%
   \hfil\raise0.36ex\hbox{$\scriptscriptstyle\boldsymbol{-}$}\hfil\cr}}}

\newcommand{\ours}{\kw{BoxDPpS}}
\newcommand{\advexact}{\kw{AdvExactGVIt}}
\newcommand{\MAvgP}{\kw{MAvgP}}
\newcommand{\iBF}{\kw{iBF}}
\newcommand{\Full}{\kw{Full}}
\newcommand{\NoPrim}{\kwnospace{No}-\kw{Prim}}
\newcommand{\NoBoxUB}{\kwnospace{No}-\kw{BoxUB}}
\newcommand{\NoWS}{\kwnospace{No}-\kw{W/S}}
\newcommand{\NoGroup}{\kwnospace{No}-\kw{Group}}

\usepackage{weiwAlgorithm}

\begin{document}
\title{Scalable Exact Densest $P$-Partite Subgraph Search in Heterogeneous Information Networks}

\author{Jiadong Xie}
\authornote{Jiadong Xie and Jiaming Yang are the joint first authors.}
\orcid{0000-0003-4535-8359}
\affiliation{%
  \institution{The Chinese University of Hong Kong}
  \country{}
}
\email{jdxie@se.cuhk.edu.hk}

\author{Jiaming Yang}
\authornotemark[1]
\affiliation{%
\institution{Beijing Institute of Technology}
\country{}
}
\email{yumoym456@gmail.com}

\author{Kangfei Zhao}
\orcid{0000-0002-7189-983X}
\affiliation{%
\institution{Beijing Institute of Technology}
\city{}
\country{}
}
\email{zkf1105@gmail.com}

\author{Jeffrey Xu Yu}
\orcid{0000-0002-9738-827X}
\affiliation{%
  \institution{HKUST (Guangzhou)}
  \country{}
}
\email{jeffreyxuyu@hkust-gz.edu.cn}

\begin{abstract}
Heterogeneous information networks (HINs) model typed entities and typed relations, where dense cross-type structures can reveal cohesive semantic patterns such as prolific author-paper-venue groups. Given a query meta-path, the densest $P$-partite subgraph search (DPpS) problem jointly selects a nonempty vertex set at each typed position and maximizes the number of induced meta-path instances normalized by the geometric mean of the selected set sizes.
Existing exact methods solve DPpS by searching over iRM-sets and reducing each fixed-$\mathbf{M}$ problem to minimum-cut computations. However, their scalability is limited by the large number of candidate iRM-sets and the high cost of repeatedly solving large auxiliary networks.
In this paper, we propose \ours, an efficient exact approach that reduces both sources of cost. It performs box-level search with safe region pruning, eliminates redundant representations of the same iRM-set, improves early pruning through bounded warm-up, and compresses each fixed-$\mathbf{M}$ auxiliary network for exact parametric pseudoflow solving.
Experiments on seven real-world datasets show that \ours preserves the exact DPpS optimum while achieving an average speedup of $27.04\times$ over the state-of-the-art method.
\end{abstract}

\maketitle

\ifdefempty{\vldbavailabilityurl}{}{
\vspace{.3cm}
\begingroup\small\noindent\raggedright\textbf{PVLDB Artifact Availability:}\\
The source code, data, and/or other artifacts have been made available at \url{\vldbavailabilityurl}.
\endgroup
}

\section{Introduction}
\label{sec:introduction}

\begin{figure}[t]
    \centering
    \resizebox{0.95\linewidth}{!}{%
        \begin{tikzpicture}[
    hinNode/.style={circle, draw, minimum size=4.8mm, inner sep=0pt, font=\tiny},
    typeA/.style={hinNode, fill=blue!12},
    typeP/.style={hinNode, fill=orange!18},
    typeV/.style={hinNode, fill=green!16},
    edge/.style={->, >=Stealth, thin, draw=gray!65},
    selected/.style={draw=red!75!black, very thick},
    selectedEdge/.style={->, >=Stealth, thick, draw=red!75!black},
    panelTitle/.style={font=\scriptsize},
    labelText/.style={font=\tiny}
]

\node[typeA] (sA) at (-0.9,1.25) {A};
\node[typeP] (sP) at (0,1.25) {P};
\node[typeV] (sV) at (0.9,1.25) {V};
\draw[edge] (sA) -- node[above, labelText] {write} (sP);
\draw[edge] (sP) -- node[above, labelText] {pubIn} (sV);
\node[labelText] at (0,0.55) {$\mathcal{P}: A \rightarrow P \rightarrow V$};

\node[typeA, minimum size=2.6mm] at (-1.20,0.05) {};
\node[labelText, anchor=west] at (-1.03,0.05) {Author};
\node[typeP, minimum size=2.6mm] at (-0.23,0.05) {};
\node[labelText, anchor=west] at (-0.06,0.05) {Paper};
\node[typeV, minimum size=2.6mm] at (0.68,0.05) {};
\node[labelText, anchor=west] at (0.85,0.05) {Venue};

\node[typeA, selected] (a1) at (3.05,0) {$a_1$};
\node[typeA, selected] (a2) at (3.65,0) {$a_2$};
\node[typeA, selected] (a3) at (4.20,0) {$a_3$};
\node[typeA, selected] (a4) at (4.75,0) {$a_4$};

\node[typeP] (p1) at (2.15,0.90) {$p_1$};
\node[typeP] (p2) at (2.65,0.90) {$p_2$};
\node[typeP] (p3) at (3.15,0.90) {$p_3$};
\node[typeP, selected] (p4) at (3.70,0.90) {$p_4$};
\node[typeP, selected] (p5) at (4.40,0.90) {$p_5$};

\node[typeV] (v1) at (2.65,1.80) {$v_1$};
\node[typeV, selected] (v2) at (3.70,1.80) {$v_2$};
\node[typeV, selected] (v3) at (4.40,1.80) {$v_3$};

\draw[edge] (a1) -- (p1);
\draw[edge] (a1) -- (p2);
\draw[edge] (a1) -- (p3);
\draw[edge] (p1) -- (v1);
\draw[edge] (p2) -- (v1);
\draw[edge] (p3) -- (v1);

\draw[selectedEdge] (a1) -- (p4);
\draw[selectedEdge] (a2) -- (p4);
\draw[selectedEdge] (a2) -- (p5);
\draw[selectedEdge] (a3) -- (p4);
\draw[selectedEdge] (a3) -- (p5);
\draw[selectedEdge] (a4) -- (p4);
\draw[selectedEdge] (a4) -- (p5);
\draw[selectedEdge] (p4) -- (v2);
\draw[selectedEdge] (p4) -- (v3);
\draw[selectedEdge] (p5) -- (v2);
\draw[selectedEdge] (p5) -- (v3);

\node[labelText, text=red!75!black] at (3.90,-0.42) {14 selected instances};
\draw[densely dashed, draw=red!60!black, rounded corners=1pt]
    (3.45,1.52) rectangle (4.65,2.08);
\node[labelText, text=red!75!black] at (4.05,2.22) {same predecessors};

\node[panelTitle] at (0,-0.72) {(a) Schema and meta-path};
\node[panelTitle] at (3.45,-0.72) {(b) Running HIN instance};

\end{tikzpicture}
    }
    \vspace{-2ex}
    \caption{An Example of Author--Paper--Venue HIN}
    \vspace{-2ex}
    \label{fig:running-example}
\end{figure}
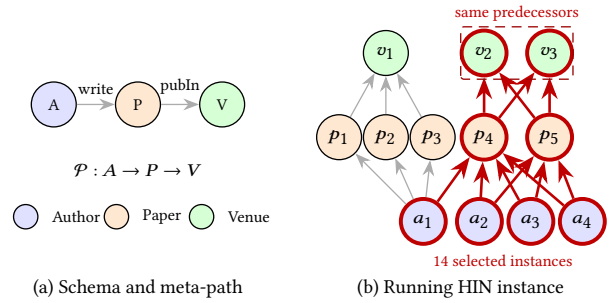

Heterogeneous information networks (HINs) model data whose entities and relations have distinct types. For example, a bibliographic network connects authors, papers, and venues, whereas a recommendation network relates users, movies, genres, and tags. 
In such networks, meaningful structures often arise from the joint participation of several entity types rather than from the connectivity of a single type. A group of authors, for instance, is better characterized together with the papers they write and the venues in which those papers are published.
Meta-paths provide a standard mechanism for specifying such cross-type semantics~\cite{SunHYYW11,ShiKHYW14,HuangZCSML16}. A meta-path describes a sequence of vertex and relation types that a valid interaction must follow. For example, the meta-path $\mathsf{Author}\rightarrow\mathsf{Paper}\rightarrow\mathsf{Venue}$ in Fig.~\ref{fig:running-example}(a) represents the semantic relation that an author writes a paper published in a venue, and each concrete author--paper--venue sequence satisfying these relations is an instance of the meta-path.

A large body of existing HIN research studies typed structures from either an object-level or a query-driven perspective. Meta-path similarity measures quantify the semantic relevance between individual objects~\cite{SunHYYW11,ShiKHYW14}, whereas HIN community-search models retrieve subgraphs satisfying query, degree, core, truss, motif, relation, or attribute constraints~\cite{FangYZLC20,DBLP:journals/pvldb/JiangFMCL22,YangF00F20,JianWC20,JiangYCHNLSL25}. These formulations are effective when an analyst has designated query vertices, a target vertex type, or a desired cohesiveness threshold. However, they are not designed to globally identify a collection of typed vertex sets that are jointly dense with respect to complete cross-type interactions.
The densest $P$-partite subgraph search (DPpS) problem~\cite{ChenLZLXL23} offers a global-density perspective. Given a query meta-path $\mathcal{P}$, DPpS jointly selects a nonempty vertex set for every typed position along $\mathcal{P}$ and maximizes the number of complete meta-path instances induced by these sets, normalized by the geometric mean of their cardinalities. 
Fig.~\ref{fig:running-example}(b) illustrates the densest vertex sets selected for an Author--Paper--Venue query.
By counting complete meta-path instances rather than individual edges or pairwise similarities, DPpS directly measures how intensively all participating types interact. Its geometric-mean normalization treats the typed positions symmetrically and discourages enlarging a selected set unless the additional vertices contribute sufficiently many complete instances. 
Thus, DPpS requires neither a query vertex nor a designated target type or cohesiveness threshold, making it suitable for exploratory discovery of compact cross-type groups. Such groups may correspond, for example, to authors publishing intensively through a collection of papers and venues, as shown in Fig.~\ref{fig:case-study-comparison}. 

It is a huge challenge to solve the DPpS problem exactly.
The DPpS objective is a ratio: its numerator depends jointly on all selected vertex sets, while its denominator is a nonlinear geometric mean whose optimal type-wise cardinalities are unknown in advance. A principled route to exactness is to combine the iRM-set formulation with a minimum-cut reduction~\cite{ChenLZLXL23}. An iRM-set $\mathbf{M}$ represents the relative cardinalities of the vertex sets selected at the different positions of $\mathcal{P}$. Once $\mathbf{M}$ is fixed, the geometric-mean denominator can be expressed as a linear weighted sum of the selected type-wise cardinalities. For a density threshold $\gamma$, determining whether a denser solution exists then becomes a problem that can be solved exactly through a minimum $s$-$t$ cut.
The state-of-the-art exact algorithm, \advexact~\cite{ChenLZLXL23}, builds on this decomposition by searching over admissible iRM-sets and invoking iterative minimum-cut computations for the corresponding fixed-$\mathbf{M}$ subproblems.
However, such methods still have two issues.
 First, since the optimal cardinalities are unknown, an exact algorithm must reason over a potentially large collection of admissible iRM-sets. The underlying count domain grows as the product of the number of available vertices at the different typed positions of the query meta-path, and even an iRM-set that is eventually pruned can incur nontrivial generation and condition-evaluation costs. Second, every surviving iRM-set requires conducting minimum-cuts, which creates one auxiliary node for every meta-path instance and connects it to its participating vertices. Because the number of meta-path instances can be much larger than the number of distinct vertices, repeatedly constructing and solving these networks is expensive.

To address the efficiency bottlenecks of \advexact, we propose \ours, an efficient exact approach that reduces both the number of iRM-sets that need to be processed and the cost of solving each surviving fixed-$\mathbf{M}$ problem. The main contributions are summarized as follows.
\ding{202}
We develop an exact box-level search to avoid exhaustively enumerating and processing the potentially large number of iRM-sets. Even if an iRM-set is eventually pruned and never reaches the fixed-$\mathbf{M}$ solver, generating it and evaluating the pruning conditions still incur substantial overhead due to point-by-point enumeration. Thus, \ours organizes count vectors into integer boxes and prunes an entire box whenever a safe upper bound proves that none of the solutions represented by the box can improve the current best density. 
\ding{203}
To reduce the number of iRM-sets processed, we eliminate redundant representations of the same iRM-set. We prove a one-to-one correspondence between the resulting canonical representations and admissible iRM-sets, ensuring that each fixed-$\mathbf{M}$ problem is processed at most once.
\ding{204}
To further increase the number of iRM-sets that can be pruned, we design a bounded warm-up strategy. It evaluates a small set of representative ratios to obtain feasible solutions and schedules the most promising ratios for early exact processing. 
\ding{205}
To accelerate each fixed-$\mathbf{M}$ problem, we compress its auxiliary network and prove that the compression preserves the original objective. We then solve the resulting compact network exactly using parametric Hochbaum pseudoflow~\cite{Hochbaum08}, reducing the per-iRM-set complexity from $O(|V(D)|^3)$ in \advexact to $O(n_M^2\log n_M)$ in \ours, where $n_M\leq |V(D)|$.
\ding{206} 
We conduct extensive experiments on seven real-world HIN datasets. On the query meta-paths completed by both exact methods, \ours achieves a speedup of $27.04\times$ over \advexact on average. It also completes several large workloads on which \advexact times out or exhausts memory, while preserving the exact DPpS optimum.

\stitle{Roadmap.} The rest of this paper is organized as follows. We introduce the preliminaries in \cref{sec:preliminary}. \cref{sec:existing-algorithm} introduces the existing algorithm and its issues. We present an overview of our algorithm \ours in \cref{sec:framework-overview}. 
\cref{sec:primitive-canonicalization}, \cref{sec:box-based-irm-search}, \cref{sec:warmup}, and \cref{sec:compact-fixed-m-solving} elaborate on the deduplication of iRM-set, box-level search, bounded warm-up, and acceleration of fixed-$\mathbf{M}$ solving, respectively. We report our experimental studies in \cref{sec:experiments} and review the related works in \cref{sec:related-work}. \cref{sec:conclusion} concludes the paper. 

\section{Preliminaries}
\label{sec:preliminary}

This section introduces the notation used throughout the paper and the densest $P$-partite subgraph search problem.
Table~\ref{tab:notations} summarizes the frequently used notations. 

\begin{table}[t]
    \centering
    \caption{Summary of Notations}
    \label{tab:notations}
    \vspace{-2ex}
    \scriptsize
    \begin{tabular}{p{0.22\columnwidth}|p{0.58\columnwidth}} \hline
        \textbf{Notation} & \textbf{Definition}\\ \hline
        $\mathcal{G}$ & input heterogeneous information network \\ \hline
        $V,E$ & vertex set and edge set of $\mathcal{G}$ \\ \hline
        $\mathcal{A},\mathcal{R}$ & vertex-type set and edge-type set \\ \hline
        $\phi,\psi$ & vertex-type and edge-type mappings \\ \hline
        $T_{\mathcal{G}}$ & network schema of $\mathcal{G}$ \\ \hline
        $\mathcal{P}=(A_1,\ldots,A_k)$ & query meta-path with $k$ typed positions \\ \hline
        $U_i$, $N_i$ & vertices admissible at position $i$ and $N_i=|U_i|$  \\ \hline
        $\mathcal{F}_{\mathcal{P}}$ & all instances of $\mathcal{P}$ in $\mathcal{G}$ \\ \hline
        $\mathcal{V}$ & $P$-family $(V_1,\ldots,V_k)$ \\ \hline
        $\mathbb{V}$ & set of all feasible $P$-families \\ \hline
        $d_i(v;\mathcal{I})$ & support of a position-$i$ vertex in an instance collection $\mathcal{I}$  \\ \hline
        $\mathbf{c}(\mathcal{V})$ & count vector $(|V_1|,\ldots,|V_k|)$ \\ \hline
        $\bar{\mathbf{c}}$ & primitive count vector obtained by gcd normalization \\ \hline
        $\mathcal{D}_{c}$ & count domain $[N_1]\times\cdots\times[N_k]$ \\ \hline
        $G(\mathbf{c})$ & geometric mean $(\prod_i c_i)^{1/k}$ \\ \hline
        $\mathbf{M}(\mathbf{c})$ & iRM-set induced by count vector $\mathbf{c}$ \\ \hline
         $\mathbb{M}$ & set of all admissible iRM-sets \\ \hline
        $\rho(\mathcal{V})$, $\rho^*$ & density of $\mathcal{V}$ and the global optimum density  \\ \hline
        $\rho_{\mathbf{M}}^*$ & optimum density restricted to families conforming to $\mathbf{M}$ \\ \hline
        $\rho_{\mathrm{best}}$ & best feasible density found so far by an algorithm \\ \hline
        $\gamma$ & density threshold used in a fixed-$\mathbf{M}$ auxiliary problem \\ \hline
    \end{tabular}
\end{table}

A \textbf{heterogeneous information network (HIN)} is a graph $\mathcal{G}=(V,E,\mathcal{A},\mathcal{R},\phi,\psi)$, where $V$ and $E$ are the vertex and edge sets, $\mathcal{A}$ and $\mathcal{R}$ are the vertex-type and edge-type sets, $\phi:V\rightarrow\mathcal{A}$ maps each vertex to a type, and $\psi:E\rightarrow\mathcal{R}$ maps each edge to a relation type. The network is heterogeneous if $|\mathcal{A}|>1$ or $|\mathcal{R}|>1$.
The \textbf{network schema} $T_{\mathcal{G}}$ is the type-level graph induced by $\mathcal{A}$ and $\mathcal{R}$. Each schema vertex represents a vertex type, and each schema edge represents an admissible relation type between two vertex types. A query over an HIN is usually specified on this schema rather than on individual vertices.

\stitle{Meta-path and Instance:}
A query meta-path $\mathcal{P}$ is a typed path on $T_{\mathcal{G}}$: $\mathcal{P}=A_1 \xrightarrow{R_1} A_2 \xrightarrow{R_2}\cdots \xrightarrow{R_{k-1}} A_k$, where $A_i\in\mathcal{A}$ and $R_i\in\mathcal{R}$.
For position $i\in[k]=\{1,\ldots,k\}$, let $U_i=\{v\in V\mid \phi(v)=A_i\}$ and $N_i=|U_i|$. 
When relation types are clear from context, we write $\mathcal{P}=(A_1,\ldots,A_k)$ and call $k$ the number of vertex-type positions in $\mathcal{P}$.
A vertex sequence $p=(v_1,\ldots,v_k)\in U_1\times\cdots\times U_k$ is an instance of $\mathcal{P}$ if every consecutive pair follows the required relation, i.e., $(v_i,v_{i+1})\in E$ and $\psi((v_i,v_{i+1}))=R_i$ for all $i<k$. We write $p_i=v_i$ and denote by $\mathcal{F}_{\mathcal{P}}$ the set of all such instances.
{For example, consider the meta-path
$\mathcal{P}=(\mathsf{Author},\mathsf{Paper},\mathsf{Venue})$
in Fig.~\ref{fig:running-example}. The vertex sequences
$(a_1,p_1,v_1)$ and $(a_2,p_4,v_2)$ are instances of
$\mathcal{P}$.}

Following the setting in existing work~\cite{ChenLZLXL23}, we assume that the query positions have pairwise distinct vertex types. Hence, the sets $U_1,\ldots, U_k$ are disjoint, and every query vertex has an unambiguous position.
For an instance collection $\mathcal{I}\subseteq\mathcal{F}_{\mathcal{P}}$, the support of a
position-$i$ vertex $v\in U_i$ is $d_i(v;\mathcal{I})=\bigl|\{p\in\mathcal{I}\mid p_i=v\}\bigr|.$

\stitle{$P$-family and Induced Instances:}
Given $\mathcal{P}$, a $P$-family is a tuple $\mathcal{V}=(V_1,\ldots,V_k)$ such that $\emptyset\ne V_i\subseteq U_i$ for every $i\in[k]$. Let $\mathbb{V}$
denote the set of all $P$-families. The instances induced by
$\mathcal{V}$ are
$\mathcal{F}(\mathcal{V})=\{p\in\mathcal{F}_{\mathcal{P}}\mid p_i\in V_i\text{ for all }i\in[k]\}.$
The induced $P$-partite subgraph consists of the selected vertices and the query-compatible edges between consecutive positions. Since every objective, pruning rule, and flow construction in this paper depends on the complete instances in $\mathcal{F}(\mathcal{V})$, we use the family and its induced instance set as the primary representation.

For $\mathcal{V}\in\mathbb{V}$, its count vector is
$\mathbf{c}(\mathcal{V})=(c_1,\ldots,c_k)$, where $c_i=|V_i|$. Let $G(\mathbf{c})=\left(\prod_{i=1}^{k}c_i\right)^{1/k}$ be the geometric mean of the position-wise counts.
The admissible count domain is $\mathcal{D}_{c}=[N_1]\times\cdots\times[N_k]$, where $[N_i]=\{1,\ldots,N_i\}$.

\stitle{Density:}
The density of a $P$-family $\mathcal{V}$ is defined as 
$$\rho(\mathcal{V})=\frac{|\mathcal{F}(\mathcal{V})|}{G(\mathbf{c}(\mathcal{V}))}=\frac{|\mathcal{F}(\mathcal{V})|}{(\prod_{i=1}^{k}|V_i|)^{1/k}}.$$
The numerator counts complete query instances rather than individual edges, while the geometric-mean denominator normalizes all query positions symmetrically.
{For example, the highlighted $P$-family in
Fig.~\ref{fig:running-example} has count vector $(4,2,2)$ and
induces 14 meta-path instances. Its density is
$\frac{14}{(4\times2\times2)^{1/3}}=\frac{14}{\sqrt[3]{16}}.$}

Based on the above definitions, we can formulate the Densest $P$-Partite Subgraph Search (DPpS) problem as follows.

\stitle{Densest $P$-Partite Subgraph Search Problem:}
Given an HIN $\mathcal{G}$ and a query meta-path $\mathcal{P}=(A_1,\ldots,A_k)$, the DPpS problem aims to find a $P$-family $\mathcal{V}^{*}\in\arg\max_{\mathcal{V}\in\mathbb{V}}\rho(\mathcal{V})$. The subgraph induced by $\mathcal{V}^{*}$ is called a densest $P$-partite subgraph, and we denote its density as $\rho^{*}=\max_{\mathcal{V}\in\mathbb{V}}\rho(\mathcal{V}).$

To explain how exact search checks whether a target density is attainable, we next introduce the iRM-set formulation from previous work~\cite{ChenLZLXL23}. An iRM-set encodes the relative position-wise cardinalities of a $P$-family. Once an iRM-set is fixed, the geometric-mean denominator can be rewritten as a linear weighted sum.

\stitle{iRM-set:} 
For a positive count vector $\mathbf{c}=(c_1,\ldots,c_k)\in\mathcal{D}_{c}$,
its induced iRM-set is $\mathbf{M}(\mathbf{c})=(m_1,\ldots,m_k),$ where $m_i=\frac{G(\mathbf{c})}{c_i}.$
The entries of iRM-set are indexed by the positions of $\mathcal{P}$, we therefore write it as an ordered vector.
A $P$-family $\mathcal{V}$ \emph{conforms to} $\mathbf{M}$ if $\mathbf{M}(\mathbf{c}(\mathcal{V}))=\mathbf{M}$. Let $\mathbb{M}=\{\mathbf{M}(\mathbf{c})\mid \mathbf{c}\in\mathcal{D}_{c}\}$ be the set of all admissible iRM-sets. {For example, the highlighted family in
Fig.~\ref{fig:running-example} has
$\mathbf{c}=(4,2,2)$. Hence,
$G(\mathbf{c})=(4\times2\times2)^{1/3}=\sqrt[3]{16}$, and its induced iRM-set is $\mathbf{M}(\mathbf{c})=\left(\frac{\sqrt[3]{16}}{4},\frac{\sqrt[3]{16}}{2},\frac{\sqrt[3]{16}}{2}\right).$}

For every $i,j\in[k]$, the definition gives $m_i c_i=G(\mathbf{c}),$ $ \frac{m_i}{m_j}=\frac{c_j}{c_i}$, $\prod_{i=1}^{k}m_i=1,$ and $G(\mathbf{c})=\frac{1}{k}\sum_{i=1}^{k}m_i c_i.$
The last identity converts the geometric-mean denominator into a linear weighted sum once $\mathbf{M}$ is fixed.
For $\mathbf{M}\in\mathbb{M}$, define the ratio-restricted optimum $\rho_{\mathbf{M}}^{*}=\max\{\rho(\mathcal{V})\mid\mathcal{V}\in\mathbb{V},\mathbf{M}(\mathbf{c}(\mathcal{V}))=\mathbf{M}\}.$
Thus, $\rho^{*}=\max_{\mathbf{M}\in\mathbb{M}}\rho_{\mathbf{M}}^{*}$.

\section{Existing Approach and Issues}
\label{sec:existing-algorithm}

The DPpS objective is difficult to optimize exactly because its denominator is a nonlinear geometric mean and the optimal cardinalities of the typed vertex sets are unknown. Existing exact search~\cite{ChenLZLXL23}, named \advexact, addresses this difficulty by combining iRM-sets with minimum-cut computations. At a high level, it first fixes a relative-cardinality pattern, checks whether a given density can be achieved under this pattern, and repeatedly increases the density until it can no longer be improved. Since the optimal relative-cardinality pattern is unknown, this process must be performed over multiple iRM-sets.

\stitle{From an iRM-Set to a Density Check:}
An iRM-set represents the relative cardinalities of the selected vertex sets. Consider the highlighted Author--Paper--Venue family in Fig.~\ref{fig:running-example}, whose count vector is $(4,2,2)$. This vector means that the selected numbers of authors, papers, and venues follow the ratio $2:1:1$. Its iRM-set assigns weights inversely proportional to these cardinalities. Therefore, papers and venues receive twice the cardinality weight of authors.
The benefit of fixing an iRM-set is that it converts the nonlinear denominator of the DPpS density into a linear weighted sum. Specifically, for every family conforming to $\mathbf{M}(\mathbf{c})=(m_1,\ldots,m_k)$, its geometric-mean denominator satisfies $G(\mathbf{c})=\frac{1}{k}\sum_{i=1}^{k}m_i|V_i|.$
Therefore, checking whether a density threshold $\gamma$ is attainable becomes a linear gain--cost problem: the selected meta-path instances provide the gain, whereas the selected vertices contribute a weighted cost proportional to $\gamma$. The resulting problem can be solved exactly by a minimum $s$-$t$ cut.
For example, the highlighted family in Fig.~\ref{fig:running-example} has a count vector $(4,2,2)$, so the paper and venue positions have twice the weight of the author position. It induces 14 instances and has density $14/\sqrt[3]{16}\approx5.56$. Hence, it demonstrates that the threshold $\gamma=5$ is attainable. The minimum-cut computation performs the same test over all possible vertex selections under the fixed iRM-set.

\stitle{Iterative Density Checking:}
For one fixed iRM-set, the exact solver starts from a feasible density threshold $\gamma$ and asks whether a denser family exists. If the minimum cut returns a family with density larger than $\gamma$, the threshold is raised to the density of that family and checked again. Otherwise, no family associated with this iRM-set can improve the current threshold, and the fixed iRM-set solve terminates.

\stitle{Search over iRM-Sets:}
A fixed iRM-set covers one relative-cardinality pattern. Since the optimal pattern is unknown, \advexact generates the iRM-sets induced by feasible count vectors and processes the candidates that cannot be pruned. For each such candidate, it may first obtain a feasible lower bound and reduce the graph, and then run the iterative density checks described above.

A completed exact solve for an iRM-set $\mathbf{M}^{a}=(m_1^{a},\ldots,m_k^{a})$ may improve the global best density and produce a point-wise certificate for pruning other iRM-sets. In particular, a remaining iRM-set $\mathbf{M}^{c}=(m_1^{c},\ldots,m_k^{c})$ can be removed when
$k\leq\sum_{i=1}^{k}\frac{m_i^{a}}{m_i^{c}}\leq B_a$.
Here, $B_a$ is computed from the output of the completed fixed-$\mathbf{M}^{a}$ solve rather than from $\mathbf{M}^{a}$ alone. If the solve returns a family whose induced iRM-set is $\mathbf{M}'=(m_1',\ldots,m_k')$, the certificate uses $B_a=\sum_{i=1}^{k}m_i^{a}/m_i'$. Alternatively, if the solve derives a certified upper bound $\widehat{\rho}_a$ on $\rho_{\mathbf{M}^{a}}^{*}$ and obtains a feasible family $\mathcal{V}'$ satisfying $\rho(\mathcal{V}')>\widehat{\rho}_a$, the certificate uses $B_a=k\rho(\mathcal{V}')/\widehat{\rho}_a$~\cite{ChenLZLXL23}.

Overall, the existing exact algorithm consists of an outer search over iRM-sets and an inner sequence of density checks for every surviving iRM-set. 
The overall cost contains three parts: $T_{\advexact}=T_{\mathrm{generation}}+T_{\mathrm{certificate}}+N_{\mathrm{solve}}T_{\mathrm{flow}},$
where $T_{\mathrm{generation}}$ is the cost of enumerating count
vectors and constructing their iRM-set representations,
$T_{\mathrm{certificate}}$ is the cost of checking the available
cross-iRM-set pruning conditions, and $N_{\mathrm{solve}}$ is the
number of candidates that reach exact fixed-$\mathbf{M}$ solving.
The dominant exact-solving term is commonly summarized as
$\Theta(|\mathbb{M}|T_{\mathrm{flow}})$ in the existing analysis, where $|\mathbb{M}|$ is the number of iRM-sets processed and $T_{\mathrm{flow}}$ is the cost of solving one fixed-$\mathbf{M}$ problem. The existing analysis gives $T_{\mathrm{flow}}=O(|V(D)|^3)$ for the flow network $D$, resulting in a worst-case total complexity of $O((n/k)^{4k})$~\cite{ChenLZLXL23}.

\stitle{Issues:}
The time complexity $\Theta(|\mathbb{M}|T_{\mathrm{flow}})$ exposes two multiplicative sources of cost: the number of candidate iRM-sets $|\mathbb{M}|$ that the search must process, and the cost $T_{\mathrm{flow}}$ of solving a single fixed-$\mathbf{M}$ subproblem. Both factors grow rapidly with the graph size and the meta-path length $k$. We thus analyze the two dimensions separately.

\noindent \underline{(1) iRM-set Search Issue.} 
The number of candidate iRM-sets can be very large, leading to the following three issues.

First, \advexact enumerates the entire count domain $[1,N_1]\times\cdots\times[1,N_k]$ point by point, maps each vector to its iRM-set, and only then evaluates the pruning conditions. As a consequence, an iRM-set must be generated before it can be discarded: even candidates that are ultimately pruned are individually generated and tested, so the search pays at least $\Omega(\prod_i N_i)$ work irrespective of how few candidates survive. The pruning is itself point-wise; a certificate obtained from one completed solve is re-tested against every remaining candidate, so in the worst case, certificate handling grows to $\Theta(|\mathbb{M}|^2)$. 

Second, since an iRM-set encodes only relative sizes, a count vector $\mathbf{c}$ and every integer multiple $g\mathbf{c}$ induce the same $\mathbf{M}$. 
For example, the count domain of Fig.~\ref{fig:running-example} contains $\mathbf{c}=(2,1,1)$ and $\mathbf{c}'=(4,2,2)=2\mathbf{c}$. 
Doubling every coordinate also doubles the geometric mean, i.e.,
$G(\mathbf{c}')=2G(\mathbf{c})$. Hence, the common factor cancels
in every ratio $m_i=G(\mathbf{c})/c_i$, and
$\mathbf{M}(\mathbf{c}')=\mathbf{M}(\mathbf{c})$, and pointwise count-vector enumeration in \advexact can repeatedly encounter the same iRM ratio. Even if duplicate iRM objects are removed by a set data structure, the enumeration and lookup overhead remain high due to the large number of candidate iRM-sets.

Third, effective pruning may become available only after sufficiently good feasible solutions or useful exact certificates have been obtained. However, \advexact initializes $\rho_{\mathrm{best}}=0$ and improves it only as iRM-sets are processed in the order induced by Cartesian enumeration. If promising ratios appear late in this order, the incumbent remains weak and few useful certificates are available during the early stages of the search, when the candidate space is still large. Consequently, a substantial portion of $\mathbb{M}$ may be generated and examined before effective pruning becomes possible.

\noindent \underline{(2) Fixed-$\mathbf{M}$ Solving Issue.}
For each surviving iRM-set, the fixed-$\mathbf{M}$ solver builds a flow network with one auxiliary node per meta-path instance and $k$  infinite-capacity arcs from that node to its endpoint vertices. In dense regions, the number of instances $\mathcal{F}_0$ is combinatorially larger than the number of distinct vertices—a block fully connected across positions induces up to $(n /k)^k$ instances over only $n$ vertices—so the network is dominated by $|\mathcal{F}_0|$, and its per-solve cost $T_\text{flow} = O(|V(D)|^3)$ is cubic in this inflated size. 

These issues motivate the design of our approach \ours in the next section. \ours keeps the same DPpS objective and the same fixed-$\mathbf{M}$ auxiliary formulation, but changes how count ratios are generated and how to represent and solve fixed-$\mathbf{M}$ closure.

\section{Overview of \ours}
\label{sec:framework-overview}

\cref{sec:existing-algorithm} shows that the cost of exact DPpS search is determined by two multiplicative factors: the number of iRM-sets processed and the cost of the repeated density checks for each surviving iRM-set. The iRM-set search side further suffers from point-wise enumeration, duplicate ratio representations, and weak pruning at the beginning of the search. \ours addresses these issues in the same order while preserving the original iRM-set formulation and exactness guarantee.

\stitle{Avoiding Point-Wise Count-Vector Enumeration:}
To avoid the cost incurred by \advexact in enumerating every count vector individually, \ours organizes the count domain into integer boxes and searches these boxes directly. A box $\mathcal{X}=\prod_{i=1}^{k}[\ell_i,h_i]$ represents all count vectors whose $i$-th coordinate lies between $\ell_i$ and $h_i$.
For example, the box $\mathcal{X}=[1,4]\times[1,2]\times[1,2]$
represents 16 Author--Paper--Venue count vectors. A point-wise search must generate these vectors before deciding whether their iRM-sets can be discarded. In contrast, for each box, \ours computes a safe upper bound $U_{\mathrm{box}}(\mathcal{X})$ on the maximum density attainable by any $P$-family whose count vector belongs to $\mathcal{X}$. If $U_{\mathrm{box}}(\mathcal{X})\leq\rho_{\mathrm{best}}$, the entire box is pruned without materializing its individual count vectors or their induced iRM-sets. Thus, a large, unpromising region can be eliminated with a single box-level test rather than by pointwise generation and evaluation.
Completed exact solves may additionally produce point-wise certificates for pruning other iRM-sets. \ours lifts such a certificate to the box level whenever it covers every count vector represented by the box, allowing all corresponding candidates to be removed together. A box that cannot be pruned is divided into smaller disjoint boxes, and only a surviving singleton is converted into its induced iRM-set and sent to fixed-$\mathbf{M}$ processing. Section~\ref{sec:box-based-irm-search} presents the safe upper bounds, certificate lifting, box-splitting procedure, and correctness proof.

\stitle{Eliminating Scale-Equivalent iRM-Sets:}
The second issue is that proportional count vectors induce the same iRM-set. For example, $(2,1,1)$ and $(4,2,2)$ describe the same Author--Paper--Venue cardinality ratio and therefore require the same fixed-$\mathbf{M}$ solve.
To avoid processing such duplicates, every surviving count vector is mapped to its primitive representative:
$\operatorname{Primitive}(\mathbf{c})=\mathbf{c}/\gcd(c_1,\ldots,c_k)$.
All vectors on the same scaling ray therefore share one canonical key. For example, $(4,2,2)$ is mapped to $(2,1,1)$, and the corresponding iRM-set is solved only once. Section~\ref{sec:primitive-canonicalization} proves that primitive count vectors are in one-to-one correspondence with admissible iRM-sets, so canonicalization removes only redundant representations and preserves every feasible relative-cardinality pattern.

\stitle{Strengthening Early iRM-Set Pruning:}
The third issue identified in Section~\ref{sec:existing-algorithm} is that effective pruning may become available late because both the initial incumbent and the set of exact certificates are weak. This issue is particularly important for the box search of \ours, because a safe box upper bound can prune a box only when it does not exceed the current best feasible density $\rho_{\mathrm{best}}$. A stronger incumbent therefore allows more unpromising boxes to be removed early.
To obtain a useful lower bound without exhaustively solving the search space, \ours performs a bounded warm-up before the main box traversal. It evaluates a small collection of representative cardinality ratios and uses the returned feasible families to improve $\rho_{\mathrm{best}}$. The ratios producing the best feasible solutions are then processed exactly as early seeds. Besides further improving $\rho_{\mathrm{best}}$, these completed exact solves may generate iRM-set certificates that become available to the subsequent box search. Approximate results affect only the feasible lower bound and the processing order; they are never used as pruning certificates.~\cref{sec:warmup} presents the candidate-selection procedure and its safety argument.

\stitle{Accelerating Fixed-\texorpdfstring{$\mathbf{M}$}{M} Density Checks:}
The final issue is that every surviving iRM-set requires repeated density checks on a flow network containing one auxiliary node per meta-path instance. \ours retains the same fixed-$\mathbf{M}$ gain--cost objective but removes repeated structures from its network representation.

\begin{figure}[t]
\centering
\includegraphics[width=\columnwidth]{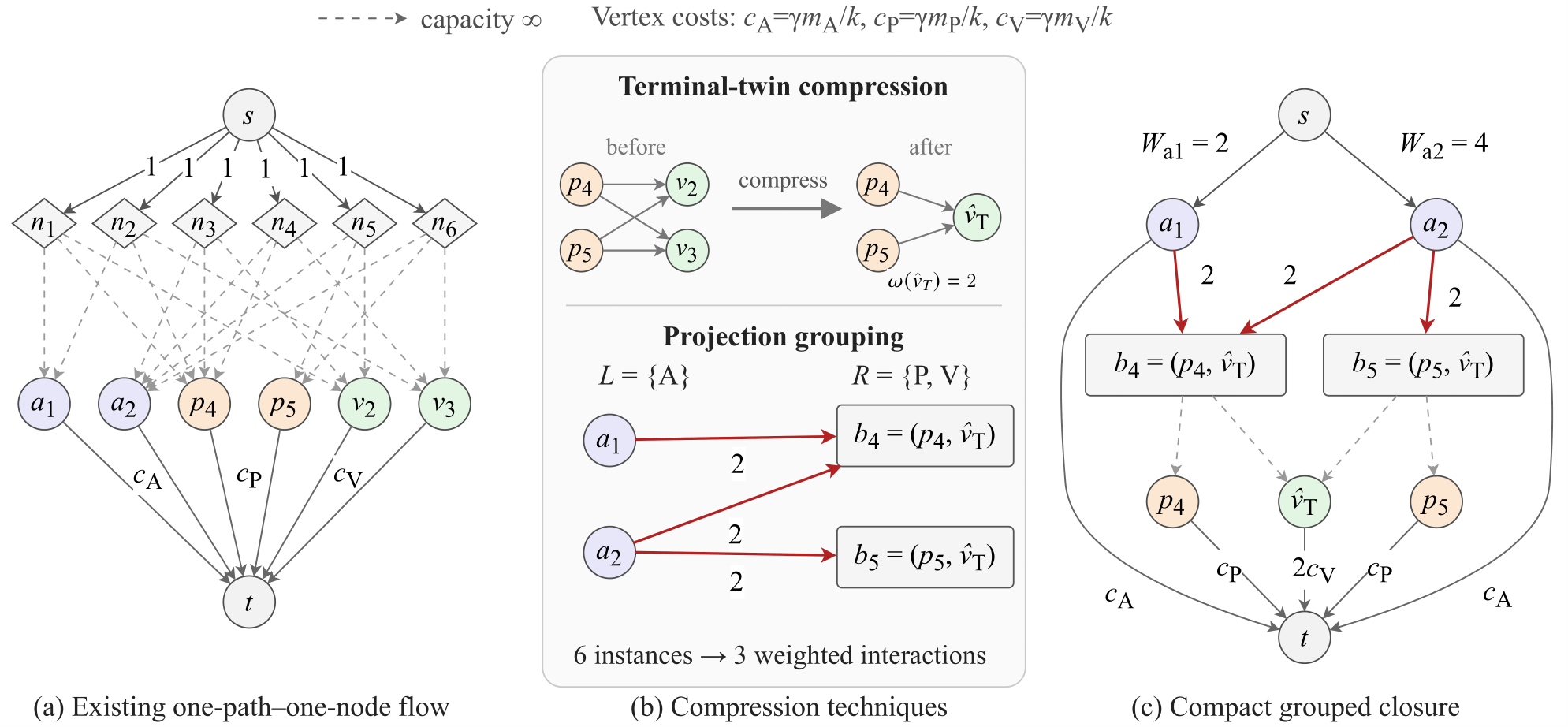}
\caption{Reducing the Network for A Fixed-$\mathbf{M}$ Density Check}
\label{fig:compress}
\end{figure}

Fig.~\ref{fig:compress} illustrates the idea using the subgraph induced by authors $\{a_1,a_2\}$, papers $\{p_4,p_5\}$, and venues $\{v_2,v_3\}$ in Fig.~\ref{fig:running-example}. This subgraph contains six Author--Paper--Venue instances, each represented separately in the existing network.
However, $v_2$ and $v_3$ have identical predecessor multisets at the terminal position in the current instance universe and can be represented by one weighted vertex. Moreover, several instances share the same projections on the selected groups of meta-path positions. They can therefore be aggregated into weighted interactions. In this example, six individual instance representations are reduced to three grouped interactions.
We prove that these transformations preserve the fixed-$\mathbf{M}$ gain--cost objective for every density threshold. The compact network therefore returns exactly the same answer as the original network, while its size depends on the number of distinct structural signatures rather than directly on the total number of meta-path instances.

Recall that a fixed-$\mathbf{M}$ solve repeatedly increases $\gamma$ and checks whether the new density can be exceeded. Across these checks, the structure of the compact network remains unchanged; only the vertex costs increase linearly with $\gamma$. We exploit this monotone structure using exact parametric Hochbaum pseudoflow~\cite{Hochbaum08}. Under the sparse compact-network conditions analyzed in Section~\ref{sec:compact-fixed-m-solving}, the per-iRM-set complexity is reduced from $O(|V(D)|^3)$ to $O(n_M^2\log n_M)$, where $n_M\leq |V(D)|$.

As a summary, Algorithm~\ref{alg:framework-overview} is the pseudocode for the overall framework of \ours. 
It first initializes the best feasible solution and the two sets used to record processed primitive keys and exact pruning certificates (lines~\ref{algo-overview-init-best}--\ref{algo-overview-init-state}). The bounded warm-up then improves the initial lower bound, after which several promising ratios are processed exactly to obtain early solutions and certificates (lines~\ref{algo-overview-warmup}--\ref{algo-overview-seeds}).
The complete count domain is initially represented by a single box (line~\ref{algo-overview-init-frontier}). During the main traversal, \ours repeatedly removes a box from the frontier (lines~\ref{algo-overview-loop}--\ref{algo-overview-pop}). If a safe upper bound or an exact certificate proves that the box cannot contain a better solution, the entire box is discarded without generating its individual count vectors (lines~\ref{algo-overview-prune-test}--\ref{algo-overview-prune}).
For a surviving singleton box, \ours maps its count vector to the corresponding primitive key (lines~\ref{algo-overview-singleton-test}--\ref{algo-overview-canonicalize}). If the key has not been processed, its fixed-$\mathbf{M}$ problem is solved using the compact parametric solver, and the resulting solution, key, and certificate are used to update the global state (lines~\ref{algo-overview-new-key}--\ref{algo-overview-update}). A surviving non-singleton box is divided into disjoint child boxes for further processing (line~\ref{algo-overview-split}).
When the frontier is exhausted, every admissible iRM-set has either been safely pruned or exactly processed through its unique primitive key. The algorithm therefore returns the globally optimal $P$-family (line~\ref{algo-overview-return}).

\begin{algorithm}[t]
  \small  
      \SetVline      \SetFuncSty{textsf}
      \SetArgSty{textsf}
   \caption{\ours($\mathcal{G},\mathcal{P}$)}
   \label{alg:framework-overview}
   \Input{HIN $\mathcal{G}$ and query meta-path $\mathcal{P}$}
   \Output{The densest $P$-family $\mathcal{V}_{\mathrm{best}}$}

Initialize the best feasible family $\mathcal{V}_{\mathrm{best}}$ and its density $\rho_{\mathrm{best}}$;
\nllabel{algo-overview-init-best}

Initialize the processed primitive keys $\mathcal{K}$ and exact certificates $\mathcal{C}$;
\nllabel{algo-overview-init-state}

Run the bounded warm-up and update $\rho_{\mathrm{best}}$ using its feasible families;
\tcp*[f]{Section~\ref{sec:warmup}}\nllabel{algo-overview-warmup}

Process the selected exact seeds and add any resulting exact certificates to $\mathcal{C}$;
\tcp*[f]{Sections~\ref{sec:box-based-irm-search}, \ref{sec:warmup}, and~\ref{sec:compact-fixed-m-solving}}\nllabel{algo-overview-seeds}

Initialize the search frontier $\mathcal{Q}$ with the box $\mathcal{X}_0$;\nllabel{algo-overview-init-frontier}

\While{$\mathcal{Q}$ is not empty\nllabel{algo-overview-loop}}{
Remove a box $\mathcal{X}$ from $\mathcal{Q}$;
\nllabel{algo-overview-pop}

\If{$\mathcal{X}$ is covered by a safe upper bound or an exact certificate\nllabel{algo-overview-prune-test}}{
    Continue;
    \tcp*[f]{Section~\ref{sec:box-based-irm-search}}\nllabel{algo-overview-prune}
}

\ElseIf{$\mathcal{X}$ is a singleton $\{\mathbf{c}\}$\nllabel{algo-overview-singleton-test}}{
    $\bar{\mathbf{c}}\leftarrow\operatorname{Primitive}(\mathbf{c})$;
    \tcp*[f]{Section~\ref{sec:primitive-canonicalization}}\nllabel{algo-overview-canonicalize}

    \If{$\bar{\mathbf{c}}\notin\mathcal{K}$\nllabel{algo-overview-new-key}}{
        Solve the fixed-$\mathbf{M}(\bar{\mathbf{c}})$ problem exactly using the compact parametric solver;
        \tcp*[f]{Section~\ref{sec:compact-fixed-m-solving}}\nllabel{algo-overview-exact-solve}

        Update $\mathcal{V}_{\mathrm{best}}$, $\rho_{\mathrm{best}}$, $\mathcal{K}$, and $\mathcal{C}$;
        \tcp*[f]{Sections~\ref{sec:box-based-irm-search} and~\ref{sec:compact-fixed-m-solving}}\nllabel{algo-overview-update}
    }
}
\Else{
    Divide $\mathcal{X}$ into disjoint child boxes and insert them into $\mathcal{Q}$;\nllabel{algo-overview-split}
}

}

\Return{$\mathcal{V}_{\mathrm{best}}$};
\nllabel{algo-overview-return}
\end{algorithm}

\section{Primitive Count Vector}
\label{sec:primitive-canonicalization}

\cref{sec:framework-overview} uses gcd normalization as the key of an iRM ratio. This section formally proves that the normalization is exact and quantifies how many duplicate count-vector representations it removes.

For a positive integer count vector $\mathbf{c}=(c_1,\ldots,c_k)$, let $g(\mathbf{c})=\gcd(c_1,\ldots,c_k)$ and $\bar{\mathbf{c}}=\mathbf{c}/g(\mathbf{c})$. We call $\bar{\mathbf{c}}$ the \emph{primitive representative} of $\mathbf{c}$ and let $\mathcal{D}_{\mathrm{prim}}=\{\mathbf{c}\in\mathcal{D}_{c}\mid g(\mathbf{c})=1\}$.
This section keeps only the two properties needed by the algorithm: canonicalization is exact, and the number of canonical keys is smaller than the absolute count domain.

We first prove that processing each primitive key once preserves the global DPpS optimum.

\begin{theorem}
\label{thm:primitive-bijection}
For positive vectors $\mathbf{x}$ and $\mathbf{y}$, $\mathbf{M}(\mathbf{x})=\mathbf{M}(\mathbf{y})$ if and only if $\mathbf{y}=\alpha\mathbf{x}$ for some $\alpha>0$.  Consequently, the mapping $\mathbf{c}\mapsto\mathbf{M}(\mathbf{c})$ is a bijection from $\mathcal{D}_{\mathrm{prim}}$ to the admissible iRM-set space $\mathbb{M}$, and $\rho^{*}=\max_{\bar{\mathbf{c}}\in\mathcal{D}_{\mathrm{prim}}}\rho^{*}_{\mathbf{M}(\bar{\mathbf{c}})}.$
\end{theorem}

\proofsketch
\sloppy
If $\mathbf{y}=\alpha\mathbf{x}$, then for every $i$, $m_i(\mathbf{y})={(\prod_j\alpha x_j)^{1/k}}/{(\alpha x_i)}=m_i(\mathbf{x}).$
Conversely, equality of the $i$-th iRM coordinates gives $y_i/x_i=G(\mathbf{y})/G(\mathbf{x})$ for every $i$, so one positive scalar relates all coordinates.  Every admissible iRM-set therefore has a primitive representative obtained by gcd normalization.  If two primitive integer
vectors induce the same iRM-set, they are positive scalar multiples.  Writing the scalar in lowest rational terms, integrality forces its denominator to divide every coordinate of the first vector, and primitivity forces that denominator to be one; primitivity of the second vector then forces the integer numerator to be one.  Hence, the primitive representative is unique. The displayed optimization identity follows because the original exhaustive search is a maximum over $\mathbb{M}$.
\eop

We next consider the time complexity when utilizing the primitive count-vector canonicalization.

\begin{theorem}
\label{thm:primitive-count}
For $k\ge2$, $|\mathbb{M}|=|\mathcal{D}_{\mathrm{prim}}|=\sum_{d=1}^{\min_iN_i}\mu(d)\prod_{i=1}^{k}\left\lfloor\frac{N_i}{d}\right\rfloor,$ where $\mu$ is the M\"obius function.
If the $N_i$ grow proportionally, then $\frac{|\mathbb{M}|}{\prod_iN_i}\longrightarrow\frac{1}{\zeta(k)}.$
Hence, canonicalization removes exactly $\prod_iN_i-|\mathbb{M}|$ duplicate absolute count vectors; asymptotically, the removed fraction is $1-1/\zeta(k)$.
\end{theorem}

\proofsketch
Use the identity $[\gcd(c_1,\ldots,c_k)=1]=\sum_{d\mid\gcd(c_1,\ldots,c_k)}\mu(d)$ and sum it over $\mathcal{D}_{c}$.  After exchanging the order of summation, a fixed $d$ contributes to exactly $\prod_i\lfloor N_i/d\rfloor$ vectors.  The first formula follows, and Theorem~\ref{thm:primitive-bijection} identifies this count with $|\mathbb{M}|$.  Dividing by $\prod_iN_i$ and taking proportional limits gives $\sum_{d\ge1}\mu(d)/d^k=1/\zeta(k)$.
\eop

{A key is normalized in $O(k\log N_{\max})$ time by repeated gcd operations (or equivalently by a precomputed prime-divisor table), where $N_{\max}=\max_{i\in[k]}N_i$. Canonicalization is performed only when a singleton or an exact seed is processed; it is not used to delete a non-singleton box, whose safe bounds remain count-specific.}

Primitive canonicalization removes duplicate fixed-$\mathbf{M}$ solves, but it does not by itself justify pruning a non-singleton count box. A box may contain count vectors from many different scaling rays, and its density bounds depend on the absolute counts represented by the box. Therefore, \ours applies canonicalization only when processing an exact seed or a surviving singleton, while all non-singleton boxes remain in the original count coordinates and are pruned only by the safe conditions developed in~\cref{sec:box-based-irm-search}.

\section{Box-Level {iRM-Set} Search}
\label{sec:box-based-irm-search}

\cref{sec:framework-overview} explains how \ours replaces point-wise iRM-set enumeration with a search over count boxes. This section formalizes the box frontier, the two box-level pruning mechanisms, and the correctness of the resulting search.

\subsection{Box Frontier and Safe Deletion}
\label{subsec:box-frontier-details}

For a box $\mathcal{X}$ as defined in~\cref{sec:framework-overview}, a box is a singleton when $\ell_i=h_i$ for every $i$.
Let $\rho_{\max}(\mathcal{X})$ be the largest DPpS density among feasible families whose count vector lies in $\mathcal{X}$.  A function $U(\mathcal{X})$ is safe when $U(\mathcal{X})\ge\rho_{\max}(\mathcal{X})$.

\begin{lemma}
\label{lem:generic-box-pruning}
If $U(\mathcal{X})$ is a safe upper bound and $U(\mathcal{X})\le\rho_{\mathrm{best}}$, removing the entire box cannot remove a solution that improves the current incumbent.
\end{lemma}

\proofsketch
Every feasible family represented by the box has density at most $U(\mathcal{X})$, which is at most the density of an already feasible family. Hence no family in the box can strictly improve the incumbent.
\eop

For a non-singleton box, define its multiplicative span at dimension $i$ as $\sigma_i(\mathcal{X})=\log(h_i+1)-\log\ell_i$. The implementation chooses $r\in\arg\max_{i:\ell_i<h_i}\sigma_i(\mathcal{X})$, breaking ties by the raw width $h_i-\ell_i$. With $q=\lfloor(\ell_r+h_r)/2\rfloor$, it creates $\mathcal{X}^{-}=\{\mathbf{c}\in\mathcal{X}:c_r\le q\},$ and $\mathcal{X}^{+}
 =\{\mathbf{c}\in\mathcal{X}:c_r\ge q+1\}.$ All other coordinate intervals are unchanged.

\begin{lemma}
\label{lem:box-split-partition}
For every non-singleton box $\mathcal{X}$, the two children $\mathcal{X}^{-}$ and $\mathcal{X}^{+}$ constructed above satisfy (i) $\mathcal{X}^{-}\cap\mathcal{X}^{+}=\emptyset$, (ii) $\mathcal{X}^{-}\cup\mathcal{X}^{+}=\mathcal{X}$, and (iii) repeated splitting terminates after finitely many operations at singleton boxes.
\end{lemma}

\proofsketch
Because $\ell_r<h_r$, the midpoint satisfies $\ell_r\le q<h_r$. Every integer $c_r\in[\ell_r,h_r]$ satisfies exactly one of $c_r\le q$ and $c_r\ge q+1$, proving disjointness and union. Each child has a strictly smaller width in dimension $r$, while no width increases. The nonnegative integer potential $\sum_i(h_i-\ell_i)$ therefore strictly decreases along every root-to-leaf path and reaches zero exactly at a singleton.
\eop

The active boxes form a disjoint frontier because the root represents the whole domain and every replacement obeys Lemma~\ref{lem:box-split-partition}. We use a stack and insert the upper child before the lower child, so the lower-count child is examined next. This order can improve the incumbent earlier but has no role in correctness. Every box is tested with the current $\rho_{\mathrm{best}}$ and all certificates available when it is popped, so later improvements automatically strengthen the tests for the remaining frontier.

\subsection{Safe Box Upper Bounds}
\label{subsec:safe-box-upper-bounds}

All box-level statistics are computed from the complete query-induced instance multiset $\mathcal{F}_0=\mathcal{F}_{\mathcal{P}}$.
An exact weighted representation may be used internally, but each stored weight equals the number of original instances it represents.
These summaries are independent of the processed box and are reused throughout the traversal.

We combine several complementary safe upper bounds.
By Lemma~\ref{lem:generic-box-pruning}, each bound can safely prune
$\mathcal{X}$ when $U(\mathcal{X})\leq\rho_{\mathrm{best}}$.

\stitle{Total-instance Bound:}
Let $\mathcal{F}_0$ be the set of all meta-path instances in the current search universe, with multiplicity included when compression is used. No family can contain more than $|\mathcal{F}_0|$ total instance weight, and every $\mathbf{c}\in\mathcal{X}$ satisfies $c_i\ge \ell_i$. Thus $U_{\mathrm{all}}(\mathcal{X})= \frac{|\mathcal{F}_0|}{(\prod_{i=1}^{k}\ell_i)^{1/k}}$ is safe.

\stitle{Degree Bound:}
Let $d_i^0(v)=d_i(v;\mathcal{F}_0)$.
Sort vertices at position $i$ by non-increasing $d_i^0(v)$ and let $D_i(t)$ be the prefix sum of the largest $t$ degrees. For any family with $|V_i|=t$, its induced instance count is at most $D_i(t)$. Therefore, $U_{\deg,i}(\mathcal{X})=\max_{\ell_i\le t\le h_i}\frac{D_i(t)}{(t\prod_{j\ne i}\ell_j)^{1/k}}$ is safe for position $i$, and $U_{\deg}(\mathcal{X})=\min_i U_{\deg,i}(\mathcal{X})$ is safe.

\stitle{Projection-aware Bound:}
For a position subset $S\subset[k]$ with $2\le |S|<k$, group instances by the projection $\pi_S(p)=(p_i)_{i\in S}$. Let $w_S(a)$ be the total weight of group $a$, and let $T_S(q)$ be the sum of the largest $q$ group weights. A family in $\mathcal{X}$ can activate at most $\prod_{i\in S}h_i$ such projection keys. Counting all paths under those keys only overestimates the numerator, so $U_{\mathrm{proj},S}(\mathcal{X})=\frac{T_S(\prod_{i\in S}h_i)}{(\prod_i\ell_i)^{1/k}}$ is safe.

\noindent\textbf{Pair-projection bound.}
For positions $(r,c)$, form a weighted matrix $A^{r,c}$ where $A^{r,c}_{uv}$ is the number of meta-path instances with $p_r=u$ and $p_c=v$. Let $H_{r,c}(a,b)=\max_{|R|\le a, |C|\le b}\sum_{u\in R}\sum_{v\in C}A^{r,c}_{uv}.$
Then $U_{\mathrm{pair},r,c}(\mathcal{X})=\frac{H_{r,c}(h_r,h_c)}{(\prod_i\ell_i)^{1/k}}$ is safe because every family in the box selects at most $h_r$ rows and $h_c$ columns at the two positions.

\begin{theorem}
\label{thm:combined-ub}
Let $U(\mathcal{X})$ be the minimum of any subset of the total-instance, degree, projection-aware, and pair-projection bounds above. Then $U(\mathcal{X})$ is a safe upper bound for $\mathcal{X}$.
\end{theorem}

\proofsketch
The total-instance bound uses the global numerator upper bound and the minimum possible denominator in the box. The degree bound follows because each induced path contributes to exactly one selected vertex at any fixed position, so the numerator is at most the sum of the largest selected-position degrees. The projection-aware bound relaxes the family to choose the heaviest possible projection keys, and the pair-projection bound further relaxes it to choose the heaviest row/column submatrix. Each relaxation can only increase the numerator and each denominator is lower-bounded by $(\prod_i\ell_i)^{1/k}$. Therefore, every listed bound is safe, and its minimum remains safe.
\eop

\subsection{Lifting Exact iRM Certificates to Boxes}
\label{subsec:box-certificate-lifting}

A completed exact fixed-$\mathbf{M}^a$ solve may generate a pointwise pruning condition of the form $k\le \sum_{i=1}^{k}\frac{m_i^a}{m_i}\le B_a,$ under the premises of the exact iRM pruning lemmas. We use such a condition only when it is backed by a strict full exact certificate. For a count vector $\mathbf{c}$, define $C_a(\mathbf{c})=\sum_{i=1}^{k}\frac{m_i^a}{m_i(\mathbf{c})}=\frac{\sum_i m_i^ac_i}{(\prod_jc_j)^{1/k}}.$
The lower bound $C_a(\mathbf{c})\ge k$ follows from AM--GM because both iRM-sets have product one.

To certify an entire box, we maximize $C_a$ over a log-box relaxation. Let $c_i=e^{y_i}$ with $y_i\in[\log\ell_i,\log h_i]$. Then $C_a(e^{\mathbf{y}})=\sum_{i=1}^{k}m_i^a\exp\left(y_i-\frac{1}{k}\sum_{j=1}^{k}y_j\right),$ which is a convex function of $\mathbf{y}$.

\begin{theorem}
\label{thm:corner-certificate}
Given a valid point-wise certificate $(\mathbf{M}^a,B_a)$, if every corner of the log-box satisfies $C_a(e^{\mathbf{y}})\le B_a$, then every integer count vector in $\mathcal{X}$ satisfies the pointwise iRM pruning condition, and the whole box can be safely pruned.
\end{theorem}

\proofsketch
The log-box is the convex hull of its $2^k$ corners. Since $C_a(e^{\mathbf{y}})$ is convex in $\mathbf{y}$, Jensen's inequality implies that its maximum over the log-box is attained at a corner. Thus all points in the box satisfy $C_a\le B_a$. The lower inequality $C_a\ge k$ holds by AM--GM. The strict exact certificate then proves that no iRM-set induced by the box can improve $\rho_{\mathrm{best}}$.
\eop

\subsection{Processing a Surviving Canonical Key}
\label{subsec:process-key}

A singleton reached by the frontier and an exact seed produced by warm-up are handled identically. Algorithm~\ref{alg:process-key} first canonicalizes the count vector, suppresses a duplicate key, invokes the exact compact fixed-$\mathbf{M}$ solver, and updates the two pieces of global state that can change: the feasible incumbent and the strict-certificate set.

\begin{algorithm}[t]
  \small  
      \SetVline      \SetFuncSty{textsf}
      \SetArgSty{textsf}
   \caption{\kw{ProcessKey}($\mathbf{c},\mathcal{G},\mathcal{P},\mathcal{K},\mathcal{C},\mathcal{V}_{\mathrm{best}},\rho_{\mathrm{best}}$)}
   \label{alg:process-key}
   \Input{Count vector $\mathbf{c}$, query $(\mathcal{G},\mathcal{P})$, and mutable state $(\mathcal{K},\mathcal{C},\mathcal{V}_{\mathrm{best}},\rho_{\mathrm{best}})$}
   \Output{Updated mutable state $(\mathcal{K},\mathcal{C},\mathcal{V}_{\mathrm{best}},\rho_{\mathrm{best}})$}
$\bar{\mathbf{c}}\leftarrow\mathbf{c}/\gcd(c_1,\ldots,c_k)$\;\nllabel{algo-process-normalize}
\State{\textbf{if} $\bar{\mathbf{c}}\in\mathcal{K}$\nllabel{algo-process-duplicate-test} \textbf{then} \Return{}}
$\mathbf{M}\leftarrow\mathbf{M}(\bar{\mathbf{c}})$\;\nllabel{algo-process-materialize-m}
$(\widehat{\mathcal{V}},\gamma_{\mathrm{final}},\mathsf{cert})\leftarrow
\kw{FixMSolve}(\mathcal{G},\mathcal{P},\mathbf{M},\rho_{\mathrm{best}})$;
\tcp*[f]{Algorithm~\ref{alg:compact-parametric-solve}}\nllabel{algo-process-compact-solve}\\
\State{\textbf{if} $\widehat{\mathcal{V}}\ne\emptyset$ and $\rho(\widehat{\mathcal{V}})>\rho_{\mathrm{best}}$\nllabel{algo-process-improve-test} \textbf{then} $\mathcal{V}_{\mathrm{best}}\leftarrow\widehat{\mathcal{V}}$;
    $\rho_{\mathrm{best}}\leftarrow\rho(\widehat{\mathcal{V}})$\nllabel{algo-process-update-incumbent}}
insert $\bar{\mathbf{c}}$ into $\mathcal{K}$\;\nllabel{algo-process-memoize}
\State{\textbf{if} $\mathsf{cert}$ is a strict full exact certificate\nllabel{algo-process-cert-test} \textbf{then} add $\mathsf{cert}$ to $\mathcal{C}$\nllabel{algo-process-add-cert}}
\end{algorithm}

The gcd normalization in line~\ref{algo-process-normalize} maps every point on a scaling ray to the unique key proved in Theorem~\ref{thm:primitive-bijection}; the membership test then prevents a second solve for that iRM-set (line~\ref{algo-process-duplicate-test}). The
iRM weights are materialized only for a new key (line~\ref{algo-process-materialize-m}), which is passed to the exact compact solver in line~\ref{algo-process-compact-solve}. The call returns the best feasible family it discovers, the final threshold of the completed exact fixed-$\mathbf{M}$ iteration, and an optional cross-ratio certificate.
Only a feasible family can raise the global incumbent (line~\ref{algo-process-improve-test}). After the exact call, the key is memoized (line~\ref{algo-process-memoize}). A certificate enters $\mathcal{C}$ only when the full exact premises of the original iRM pruning result hold (line~\ref{algo-process-cert-test}); otherwise $\mathsf{cert}=\bot$ and no cross-ratio deletion is enabled. 

\subsection{Correctness of the Box Search}

\begin{theorem}
\label{thm:box-search-correctness}
Assume that every box deletion uses either a safe upper bound or a strict full exact iRM certificate lifted by Theorem~\ref{thm:corner-certificate}, and that every surviving singleton is normalized according to Theorem~\ref{thm:primitive-bijection} and processed by Algorithm~\ref{alg:process-key}. Then the box search preserves the global DPpS optimum.
\end{theorem}

\proofsketch
Let $\mathcal{V}^{*}$ be a globally optimal $P$-family, with count vector $\mathbf{c}^{*}=\mathbf{c}(\mathcal{V}^{*})$. Consider the unique chain of frontier boxes containing $\mathbf{c}^{*}$. While the incumbent is strictly smaller than $\rho^{*}$, no box on this chain can be removed by a safe upper bound, because that bound must be at least
$\rho(\mathcal{V}^{*})=\rho^{*}$. Nor can the chain be removed by a lifted strict full exact certificate, because such a certificate proves that every represented iRM-set cannot improve the incumbent, whereas $\mathbf{M}(\mathbf{c}^{*})$ can improve it. Hence, unless the incumbent is
already optimal, the chain is divided down to the singleton $\{\mathbf{c}^{*}\}$.

At that singleton, Algorithm~\ref{alg:process-key} computes the canonical key $\bar{\mathbf{c}}^{*}$. If the key is new, Algorithm~\ref{alg:compact-parametric-solve} solves its fixed-$\mathbf{M}$ problem exactly and therefore considers $\mathcal{V}^{*}$ among its feasible families. If the key is already in $\mathcal{K}$, the same iRM-set was exactly processed earlier from another point on the scaling ray. In either case, the incumbent has already reached $\rho^{*}$ or is raised to $\rho^{*}$. Therefore, the box search cannot lose the global optimum.
\eop

\section{Representative-Ratio Warm-Up}
\label{sec:warmup}

\cref{sec:framework-overview} uses warm-up only to obtain an early feasible lower bound and schedule a few exact keys.  This section specifies that bounded procedure.  Let $\mathbf{1}=(1,\ldots,1)$ and $\mathbf{N}=(N_1,\ldots,N_k)$.  For position $i$, let $\mathbf{a}_i(q)$ have coordinate $q$ at $i$ and one elsewhere, and define $Q_i=\left\{N_i,\max\{1,\lfloor N_i/2\rfloor\},\max\{1,\lfloor N_i/4\rfloor\},\max\{1,\operatorname{round}(\sqrt{N_i})\}\right\}.$
The budget $B_w$ bounds approximate evaluations, and $B_e$ bounds exact seeds.

\begin{algorithm}[t]
  \small  
      \SetVline      \SetFuncSty{textsf}
      \SetArgSty{textsf}
   \caption{\kw{Warm-Up}($\mathcal{G},\mathcal{P},\mathcal{X}_0,B_w,B_e$)}
   \label{alg:warmup}
   \Input{HIN $\mathcal{G}$, meta-path $\mathcal{P}$, root box $\mathcal{X}_0$, and budgets $B_w,B_e$}
   \Output{Warm-up records $\mathcal{R}$ and ordered exact-seed keys $\mathcal{S}$}
$\mathcal{B}\leftarrow
\{\mathbf{1},\mathbf{N}\}\cup
\{\mathbf{a}_i(q):q\in Q_i,\ i\in[k]\}$\;\nllabel{algo-warmup-build-representatives}
$\mathcal{Q}_w\leftarrow$ an empty FIFO queue;
$\mathcal{H}\leftarrow\emptyset$;
$\mathcal{R}\leftarrow[\,]$\;\nllabel{algo-warmup-init-state}
$\mathcal{I}_{\mathrm{leaf}}\leftarrow$ a lazy lower-child-first singleton iterator over $\mathcal{X}_0$\;\nllabel{algo-warmup-init-leaf-iterator}
\ForEach{$\mathbf{c}\in\mathcal{B}$ in the displayed order\nllabel{algo-warmup-seed-loop}}{
    $\bar{\mathbf{c}}\leftarrow\operatorname{Primitive}(\mathbf{c})$\;\nllabel{algo-warmup-seed-normalize}
    \State{\textbf{if} $\bar{\mathbf{c}}\notin\mathcal{H}$\nllabel{algo-warmup-seed-unseen-test} \textbf{then} insert $\bar{\mathbf{c}}$ into $\mathcal{H}$ and enqueue it into $\mathcal{Q}_w$\nllabel{algo-warmup-seed-enqueue}}
}
$j\leftarrow0$\;\nllabel{algo-warmup-init-counter}
\While{$j<B_w$\nllabel{algo-warmup-budget-loop}}{
    \While{$\mathcal{Q}_w=\emptyset$ and $\mathcal{I}_{\mathrm{leaf}}$ has a next singleton\nllabel{algo-warmup-fallback-loop}}{
        $\mathbf{c}'\leftarrow\operatorname{next}(\mathcal{I}_{\mathrm{leaf}})$;
        $\bar{\mathbf{c}}'\leftarrow\operatorname{Primitive}(\mathbf{c}')$\;\nllabel{algo-warmup-fallback-key}
        \State{\textbf{if} $\bar{\mathbf{c}}'\notin\mathcal{H}$\nllabel{algo-warmup-fallback-unseen-test} \textbf{then} insert $\bar{\mathbf{c}}'$ into $\mathcal{H}$ and enqueue it into $\mathcal{Q}_w$\nllabel{algo-warmup-fallback-enqueue}}
    }
    \State{\textbf{if} $\mathcal{Q}_w=\emptyset$\nllabel{algo-warmup-empty-test} \textbf{then} break}
    $\bar{\mathbf{c}}\leftarrow\operatorname{dequeue}(\mathcal{Q}_w)$;\nllabel{algo-warmup-dequeue}
    $s\leftarrow0$\;\nllabel{algo-warmup-score-init}
    {Run the approximation algorithm in~\cite{ChenLZLXL23} on} $\mathbf{M}(\bar{\mathbf{c}})$ and obtain $\widetilde{\mathcal{V}}$\;\nllabel{algo-warmup-peeling}
    
    \If{$\widetilde{\mathcal{V}}\ne\emptyset$\nllabel{algo-warmup-feasible-test}}{
        $s\leftarrow\rho(\widetilde{\mathcal{V}})$;
        $\widehat{\mathbf{c}}\leftarrow\mathbf{c}(\widetilde{\mathcal{V}})$;\nllabel{algo-warmup-witness}
        $\mathcal{A}\leftarrow\{\widehat{\mathbf{c}}\}$\;\nllabel{algo-warmup-adaptive-init}
        \For{$i\leftarrow1$ \KwTo $k$\nllabel{algo-warmup-perturb-loop}}{
            $\mathbf{c}^{-}\leftarrow\widehat{\mathbf{c}}$;
            $c_i^{-}\leftarrow\max\{1,\lfloor0.8\widehat c_i\rfloor\}$\;\nllabel{algo-warmup-minus}
            $\mathbf{c}^{+}\leftarrow\widehat{\mathbf{c}}$;
            $c_i^{+}\leftarrow\min\{N_i,\lceil1.25\widehat c_i\rceil\}$\;\nllabel{algo-warmup-plus}
            insert $\mathbf{c}^{-}$ and $\mathbf{c}^{+}$ into $\mathcal{A}$\;\nllabel{algo-warmup-add-perturbations}
        }
        \ForEach{$\mathbf{c}'\in\mathcal{A}$\nllabel{algo-warmup-adaptive-loop}}{
            $\bar{\mathbf{c}}'\leftarrow\operatorname{Primitive}(\mathbf{c}')$\;\nllabel{algo-warmup-adaptive-normalize}
            \State{\textbf{if} $\bar{\mathbf{c}}'\notin\mathcal{H}$\nllabel{algo-warmup-adaptive-unseen-test} \textbf{then} insert $\bar{\mathbf{c}}'$ into $\mathcal{H}$ and enqueue it into $\mathcal{Q}_w$\nllabel{algo-warmup-adaptive-enqueue}}
        }
    }
    append $(\bar{\mathbf{c}},\widetilde{\mathcal{V}},s,j)$ to $\mathcal{R}$;
    $j\leftarrow j+1$\;\nllabel{algo-warmup-record}
}
sort $\mathcal{R}$ by non-increasing $s$\;\nllabel{algo-warmup-sort}
$\mathcal{S}\leftarrow$ the keys in the first $\min\{B_e,|\mathcal{R}|\}$ records\;\nllabel{algo-warmup-select-seeds}
\Return{$(\mathcal{R},\mathcal{S})$}\;\nllabel{algo-warmup-return}
\end{algorithm}

Algorithm~\ref{alg:warmup} shows the details.
The deterministic representatives in line~\ref{algo-warmup-build-representatives} cover balanced, full-size, and axis-dominant regimes.  Lines~\ref{algo-warmup-init-state}--\ref{algo-warmup-seed-enqueue} initialize the queue and remove proportional duplicates before they consume budget.  If these candidates and their adaptive descendants are exhausted, the lower-child-first singleton iterator supplies unseen primitive keys without invoking the
certificate-based cross-iRM-set pruning rule (lines~\ref{algo-warmup-fallback-loop}--\ref{algo-warmup-fallback-enqueue}); line~\ref{algo-warmup-empty-test} stops when no key remains.
Each dequeued key is evaluated by the existing peeling approximation (lines~\ref{algo-warmup-dequeue}--\ref{algo-warmup-peeling}).  A returned witness is scored by its actual feasible density and exposes an observed count vector (lines~\ref{algo-warmup-score-init}--\ref{algo-warmup-witness}).  The observed vector and clipped one-coordinate $0.8$ contractions and $1.25$ expansions form the adaptive neighborhood (lines~\ref{algo-warmup-adaptive-init}--\ref{algo-warmup-add-perturbations}); they are again canonicalized and deduplicated before enqueueing (lines~\ref{algo-warmup-adaptive-loop}--\ref{algo-warmup-adaptive-enqueue}).  Finally, records are ranked by feasible density and the first $B_e$ keys become exact seeds
(lines~\ref{algo-warmup-sort}--\ref{algo-warmup-return}).

\begin{theorem}
\label{theo:warmup-safety}
Algorithm~\ref{alg:warmup} does not affect exactness provided that an approximate witness updates $\rho_{\mathrm{best}}$ only through its feasible density, the approximate output creates no certificate, and every selected seed is passed to Algorithm~\ref{alg:process-key}.
\end{theorem}

\proofsketch
A feasible witness is a lower bound on $\rho^{*}$, so replacing the incumbent by its density cannot make any safe upper-bound comparison unsound.  Candidate ranking changes only processing order.  Since approximation emits no certificate, every deletion is still justified by the same safe bound or strict exact certificate as without warm-up.  Exact seeds are ordinary canonical keys processed earlier, and memoization merely prevents a duplicate solve later.
\eop

\section{Compact Fixed-\texorpdfstring{$\mathbf{M}$}{M} Solving}
\label{sec:compact-fixed-m-solving}

\cref{sec:framework-overview} outlined the multiplicity-aware solver.
This section gives its algorithm and proves only the two representation equivalences and the HPF bound needed by the global correctness argument.  The call receives the feasible threshold $\gamma_0=\rho_{\mathrm{best}}$, used only as a pivot or warm start.  Let $W_{\mathrm{all}}$ be the total aggregated path weight; $U_M=W_{\mathrm{all}}$ is a safe upper endpoint because every nonempty family has a geometric-mean denominator at least one.

\begin{algorithm}[t]
  \small  
      \SetVline      \SetFuncSty{textsf}
      \SetArgSty{textsf}
   \caption{\kw{FixMSolve}($\mathcal{G},\mathcal{P},\mathbf{M},\gamma_0$)}
   \label{alg:compact-parametric-solve}
   \Input{HIN $\mathcal{G}$, meta-path $\mathcal{P}$, iRM-set $\mathbf{M}$, and certified initial threshold $\gamma_0$}
   \Output{Best feasible family $\widehat{\mathcal{V}}$, final threshold $\gamma_{\mathrm{final}}$, and optional strict certificate $\mathsf{cert}$}
$\mathcal{F}_{\omega}\leftarrow$ replace every verified terminal-twin class by one weighted representative\;\nllabel{algo-compact-terminal-compress}
$(L,R)\leftarrow\arg\min$ over admissible nonempty bipartitions of the projection-size estimate\;\nllabel{algo-compact-choose-partition}
$\{w_{ab}\},\{W_a\},W_{\mathrm{all}}\leftarrow$ aggregate $\mathcal{F}_{\omega}$ by $(a,b)=(\pi_L(p),\pi_R(p))$\;\nllabel{algo-compact-group-paths}
Construct the grouped closure $D_{\mathbf{M}}(\gamma)$ with capacities $W_a$, $w_{ab}$, implication arcs, and $c_v(\gamma)=\omega(v)m_{\operatorname{pos}(v)}\gamma/k$\;\nllabel{algo-compact-build-network}
$U_M\leftarrow W_{\mathrm{all}}$\;\nllabel{algo-compact-upper-endpoint}
Invoke the exact parametric HPF algorithm~\cite{Hochbaum08} on $D_{\mathbf{M}}(\gamma)$ for $\gamma\in[0,U_M]$, using $\gamma_0$ only as a pivot, and store the results in $\mathcal{B}_{\mathbf{M}}$\;\nllabel{algo-compact-hpf}
$\widehat{\mathcal{V}}\leftarrow\emptyset$;
$\gamma\leftarrow\gamma_0$\;\nllabel{algo-compact-init-iteration}
\While{$\gamma\le U_M$\nllabel{algo-compact-iteration-loop}}{
    $S\leftarrow$ the canonical source set in $\mathcal{B}_{\mathbf{M}}$ that is optimal at $\gamma$\;\nllabel{algo-compact-lookup-source}
    $\mathcal{V}_S\leftarrow$ recover the original weighted $P$-family represented by $S$\;\nllabel{algo-compact-recover-family}
    $z\leftarrow |\mathcal{F}(\mathcal{V}_S)|-(\gamma/k)\sum_i m_i|V_i^S|$\;\nllabel{algo-compact-aux-value}
    \State{\textbf{if} $\mathcal{V}_S=\emptyset$ or $z\le0$\nllabel{algo-compact-stop-test} \textbf{then} break}
    \State{\textbf{if} $\widehat{\mathcal{V}}=\emptyset$ or $\rho(\mathcal{V}_S)>\rho(\widehat{\mathcal{V}})$\nllabel{algo-compact-local-improve-test} \textbf{then} $\widehat{\mathcal{V}}\leftarrow\mathcal{V}_S$\nllabel{algo-compact-update-local-best}}
    $\gamma\leftarrow\rho(\mathcal{V}_S)$\;\nllabel{algo-compact-update-threshold}
}
$\gamma_{\mathrm{final}}\leftarrow\gamma$\;\nllabel{algo-compact-final-threshold}
$\mathsf{cert}\leftarrow$ the strict full exact iRM certificate implied by the completed iteration, or $\bot$ when its premises do not hold\;\nllabel{algo-compact-derive-certificate}
\Return{$(\widehat{\mathcal{V}},\gamma_{\mathrm{final}},\mathsf{cert})$}\;\nllabel{algo-compact-return}
\end{algorithm}

\stitle{Overall Procedure:}
Algorithm~\ref{alg:compact-parametric-solve} shows the overall procedure.
Line~\ref{algo-compact-terminal-compress} replaces only terminal-twin classes that satisfy the exact condition of Theorem~\ref{thm:terminal-compression-exact}. It then selects a projection bipartition by an estimated compact-network size (line~\ref{algo-compact-choose-partition}) and merges all instances with the same endpoint projections into one weighted pair (line~\ref{algo-compact-group-paths}). These steps change representation and multiplicity, not the objective.
The grouped network is built once with constant path-profit and implication capacities and linear vertex-to-sink capacities (line~\ref{algo-compact-build-network}). The total path weight gives a safe parameter endpoint (line~\ref{algo-compact-upper-endpoint}). The exact HPF call materializes the nested source-set representation over the certified range and uses $\gamma_0$ only as a pivot (line~\ref{algo-compact-hpf}); hence, the incumbent can guide execution without removing a parameter regime needed by the completed exact solve.
The routine starts from the certified feasible threshold (line~\ref{algo-compact-init-iteration}). At every iteration, it retrieves the source set that minimizes the grouped cut at the current threshold, recovers the represented family, and evaluates the corresponding auxiliary objective (lines~\ref{algo-compact-iteration-loop}--\ref{algo-compact-aux-value}). A nonpositive value proves that no conforming family can improve the current threshold, so the iteration stops (line~\ref{algo-compact-stop-test}). Otherwise, the feasible witness may replace the local best family and strictly raises $\gamma$ to its actual density (lines~\ref{algo-compact-local-improve-test}--\ref{algo-compact-update-threshold}). The final threshold is recorded after termination (line~\ref{algo-compact-final-threshold}). A cross-ratio certificate is emitted only if the full exact premises of the existing iRM pruning result hold (line~\ref{algo-compact-derive-certificate}); otherwise, it is $\bot$. The completed result is returned to Algorithm~\ref{alg:process-key} (line~\ref{algo-compact-return}).

\subsection{Terminal-Twin Compression}

At terminal position $k$, vertices $u$ and $v$ are terminal twins when they have the same predecessor multiset at position $k-1$ and occur only at the terminal position.  A class $T$ is represented by one node with multiplicity $\omega(\hat v_T)=|T|$.

\begin{theorem}
\label{thm:terminal-compression-exact}
Replacing every terminal-twin class by one representative whose vertex cost and incident path multiplicities are multiplied by its class size preserves both the global DPpS optimum and every fixed-$\mathbf{M}$ auxiliary optimum.
\end{theorem}

\proofsketch
Fix all selections outside one class $T$ and let $q$ be the number of selected twins.  Since all twins complete the same predecessor multiset, the induced path contribution is affine in $q$, say $A+Bq$.  The fixed-$\mathbf{M}$ vertex penalty is also affine, so the auxiliary objective is affine and has an optimum at $q=0$ or $q=|T|$.
For the density objective, let $b$ be the number of already selected non-twin terminal vertices and let $C$ be the product of the fixed counts at the other positions.  If $b>0$, the density as a continuous function of $q$ is $f(q)=\frac{A+Bq}{(C(b+q))^{1/k}}.$
After removing the positive common factor, the sign of $f'(q)$ is the sign of $B(b+q)-(A+Bq)/k$, an affine nondecreasing function of $q$.  Thus $f$ can be monotone or decrease and then increase, but cannot have a strict interior maximum; an endpoint is optimal.  If $b=0$, feasibility implies $A=0$, and $f(q)$ is proportional to $q^{1-1/k}$ for $q\ge1$, so selecting all twins is optimal.  Hence, some global optimum is also all-or-none on every class.
The weighted representative has exactly these two states.  Its multiplied path gains and vertex cost equal those of selecting all original twins, while its unselected state equals selecting none.  Applying the argument class by class preserves both optimum values.
\eop

\subsection{Projection-Grouped Closure Network}

After terminal compression, choose a nonempty bipartition $L\cup R=[k]$. For an instance $p$, let $a=\pi_L(p)$ and $b=\pi_R(p)$; all instances with the same pair are aggregated into weight $w_{ab}$, and $W_a=\sum_bw_{ab}$.  The implementation chooses the admissible bipartition with the smallest estimated grouped-network size; this changes representation only.

The grouped closure contains source $s$, sink $t$, compressed vertex nodes, and one node for each distinct left and right projection.  It has arcs $s\to a$ of capacity $W_a$, arcs $a\to b$ of capacity $w_{ab}$, infinite-capacity implications from each projection node to its member vertices, and vertex-to-sink arcs of capacity $\omega(v)\gamma m_i/k$.  A singleton-side projection node may be reused as its vertex node.

\begin{theorem}
\label{thm:grouped-flow-equivalence}
For any fixed $\mathbf{M}$ and threshold $\gamma$, the minimum cut in the projection-grouped closure maximizes exactly the same auxiliary objective $\zeta(\mathcal{V},\gamma,\mathbf{M})$ as the standard one-path-node network.
\end{theorem}

\proofsketch
Fix a set $S$ of selected original vertices.  Infinite-capacity implications allow a projection node on the source side only when all of its member vertices are in $S$.  For a compatible left projection $a$, placing $a$ on the source side cuts only weights leading to incompatible right projections, which costs no more than cutting $s\to a$ with capacity $W_a$; therefore an optimal finite cut places every compatible left projection on the source side.  The same argument applies to right projections.  Consequently, the only path weights not cut are those whose two projections are compatible with $S$, and the path-related cut cost is $W_{\mathrm{all}}-\sum_{a\subseteq S,\,b\subseteq S}w_{ab}.$
The vertex-to-sink arcs add $(\gamma/k)\sum_i m_i|V_i|$, with terminal multiplicities already preserved by Theorem~\ref{thm:terminal-compression-exact}.  Hence the complete cut value is $W_{\mathrm{all}}-\zeta(\mathcal{V},\gamma,\mathbf{M})$, exactly as in the standard construction.  Minimizing the cut therefore maximizes the same auxiliary objective.
\eop

\stitle{Network Size:}
If $n_c$ is the number of compressed vertices, $g_L$ and $g_R$ are the numbers of distinct left and right projections, and $m_{LR}$ is the number of nonzero projection pairs, then the grouped network has $O(n_c+g_L+g_R)$ nodes and $O(n_c+k g_L+k g_R+m_{LR})$ arcs. In contrast, the standard network has $|\mathcal{F}_0|$ path nodes and $\Theta(k|\mathcal{F}_0|)$ path-related
arcs. The grouped construction therefore depends on distinct signatures rather than total path multiplicity.

\subsection{HPF-Compatible Parametric Solving}

For fixed $\mathbf{M}$, all path-profit and implication capacities in the grouped closure are constant. Only vertex-to-sink capacities vary with the density threshold: $c_v(\gamma)=\omega(v)\frac{m_{\operatorname{pos}(v)}}{k}\gamma.$
These capacities are linear and monotone in $\gamma$. Therefore, the compact closure satisfies the monotone parametric cut condition required by HPF~\cite{Hochbaum08}.

\begin{theorem}
\label{thm:hpf-complexity}
Let $n_M$ and $m_M$ be the number of nodes and arcs in the grouped closure network for a fixed $\mathbf{M}$, and let $b_M$ be the number of materialized breakpoints. A fully-parametric HPF backend solves the monotone fixed-$\mathbf{M}$ closure in $O\!\left(m_Mn_M\log\!\left(2+\frac{n_M^2}{m_M}\right)+b_Mn_M\right).$
If the grouped closure is sparse, $m_M=O(n_M)$, and $b_M\le n_M$, this becomes $O(n_M^2\log n_M)$.
\end{theorem}

\proofsketch
The monotonicity condition follows from the linear nondecreasing vertex-to-sink capacities above. Hochbaum's pseudoflow analysis gives a parametric cut bound of the same asymptotic core order as one pseudoflow run; with the dynamic-tree refinement this is $O(m_Mn_M\log(2+n_M^2/m_M))$. If the implementation explicitly materializes $b_M$ breakpoint cuts and scans $n_M$ nodes for each, an output cost $O(b_Mn_M)$ is added. In a monotone parametric cut, canonical source sets are nested, and each strict breakpoint changes at least one node side, so $b_M\le n_M$. Substituting $m_M=O(n_M)$ yields $O(n_M^2\log n_M)$.
\eop

By comparison, the original DPpS analysis uses the cubic per-$\mathbf{M}$ bound $O(|V(D)|^3)$ for its explicit flow network $D$. Thus, under the sparse grouped-closure conditions of Theorem~\ref{thm:hpf-complexity}, the theoretical dependence on the compact network becomes near quadratic with one logarithmic factor, in addition to the reduction from $|V(D)|$ to $n_M$. This statement does not imply that every HPF binary is faster in practice; library binding, memory copying, and breakpoint materialization can dominate on small or moderately sized networks.

\section{Experimental Studies}
\label{sec:experiments}

We evaluate \ours from four perspectives: \ding{202} Compare \ours end-to-end efficiency with \advexact; \ding{203} Evaluate the returned subgraphs under the DPpS density objective and in a cybersecurity user-grouping application; \ding{204} Quantify the contribution of each proposed technique through an ablation study; \ding{205} Examine scalability with respect to both data size and meta-path length. We first introduce the datasets, query workloads, compared methods, and experimental environment.
\subsection{Experimental Setup}

\stitle{Datasets.}
We use seven real-world HIN datasets from different application
domains: MovieLens, DBLP, and Douban~\cite{HIN-Datasets},
DBpedia and Freebase~\cite{YangF00F20}, and Cisco g21 and
Cisco g22~\cite{MadaniAG22}. The first five datasets are used for 
the general performance evaluation, while Cisco g21 and g22 contain ground-truth user groups for the cybersecurity evaluation. 
Table~\ref{tab:datasets} summarizes the numbers of vertex types, relation types, vertices, and edges in each dataset.

\begin{table}[t]
    \centering
    \caption{Dataset Statistics}
    \label{tab:datasets}
    \vspace{-2ex}
    \footnotesize
    \begin{tabular}{@{}l|r|r|r|r@{}}
        \hline
        Dataset (Abbr. Name) & $|\mathcal{A}|$ & $|\mathcal{R}|$ & $|V|$ & $|E|$ \\
        \hline
        MovieLens (ML) & 5 & 4 & 2,672 & 104,747 \\ \hline
        DBLP      & 5 & 4 & 37,795 & 174,851 \\ \hline
        Douban (DBN)    & 6 & 6 & 37,595 & 1,714,941 \\ \hline
        DBpedia (DBP)  & 417 & 634 & 4,767,652 & 12,160,769 \\ \hline
        Freebase (FB) & 21,218 & 1,152 & 36,791,318 & 94,846,085 \\ \hline
        Cisco g21 (G21) & 4 & 3 & 2,405 & 7,704 \\ \hline
        Cisco g22 (G22) & 4 & 3 & 47,490 & 110,730 \\
        \hline
    \end{tabular}
\end{table}

\stitle{Query meta-paths.}
For each dataset and each meta-path length considered in an experiment, we generate candidate meta-paths with distinct vertex-type positions. We use 20 query meta-paths when at least 20 candidates are available and retain all candidates otherwise. 

\stitle{Compared Algorithms.}
We compare our exact framework, \ours, with \advexact~\cite{ChenLZLXL23}, whose source code implementation is provided by the authors.
For the solution-quality evaluation, we reimplement \MAvgP and \iBF
following the multipartite baseline formulations~\cite{ChenLZLXL23}. \MAvgP maximizes the average number of meta-path instances, whereas \iBF adapts the butterfly-core model~\cite{DongHYZX21} to a query meta-path. 
Each algorithm is run five times on every selected query, and we report the average running time. All compared methods are evaluated using the same queries.

\stitle{Environment.}
All algorithms are implemented in C++17 and compiled using g++ 11.4.0
with the \texttt{-O3} optimization flag. Experiments are conducted in
single-thread mode on an Ubuntu 22.04.5 server equipped with two AMD
EPYC 9554 processors and 512 GB of main memory. 
We exclude only graph loading from reported running time; all algorithm-specific preprocessing and solving costs are included. A run that exceeds 48-hour limit is marked as OOT, whereas a run terminated by memory exhaustion is marked as OOM.

\stitle{Parameter Settings.}
For the iRM-set pruning rules in Section~\ref{sec:existing-algorithm},
we use $10^{-4}$ as the certificate margin
and retain a certificate only when $B_a-k>10^{-4}$. For the
representative-ratio warm-up in Section~\ref{sec:warmup}, we set
$B_w=64$ and use $B_e=4$ by default, increasing it to $8$ when
$\prod_{i=1}^{k}N_i\ge10^8$ or $\max_iN_i\ge5000$.

\subsection{Experimental Results}

\begin{figure}[t]
    \centering
    \includegraphics[width=\columnwidth]{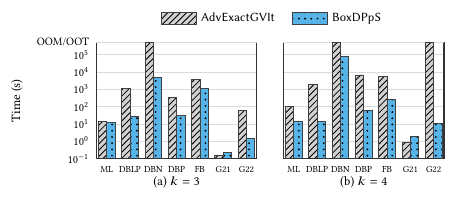}
    \Description{Two logarithmic grouped bar charts compare the running
    times of AdvExactGVIt and BoxDPpS on MovieLens, DBLP, Douban, DBpedia,
    Freebase, Cisco g21, and Cisco g22 for meta-paths with three and four
    vertex-type positions. Exact does not complete the Douban settings
    and runs out of memory on the Cisco g22 four-position setting.}
    
    \caption{Running Time of \ours and \advexact}
    \label{fig:efficiency-runtime}
\end{figure}

\stitle{Exp-1: Overall Efficiency.}
We compare the end-to-end running time of \ours and
\advexact on all seven datasets with $k=3$ and $k=4$.
Fig.~\ref{fig:efficiency-runtime} reports the running-time results.
\ours substantially outperforms \advexact on larger graphs.
Among the query meta-paths completed by both algorithms, \ours achieves an average speedup of $27.04\times$. 
On DBLP, DBpedia, and Freebase, the speedups range from $3.59\times$ to $40.74\times$ for $k=3$ and from $22.61\times$ to $131.77\times$ for $k=4$. 
Moreover, \advexact times out on both Douban settings and runs out of memory on Cisco g22 when $k=4$, whereas
\ours completes all these workloads. 
On the small Cisco g21 graph, \advexact is faster because the additional search and preprocessing
overhead of \ours is not fully amortized. Overall, the advantage becomes more remarkable as the query workload grows, demonstrating the effectiveness of box-based search and compact fixed-$\mathbf{M}$ solving.

\stitle{Exp-2: Effectiveness of Normalized Density Gain.}
We next evaluate the densities of the subgraphs returned by the compared methods. Since absolute densities are not directly comparable across query meta-paths, we measure the density gain over the complete query-induced $P$-partite graph. 
For an algorithm $a$ and a query
meta-path $p$, let $\rho_{a,p}$ denote the density of the  returned subgraph, and let $\rho_{\mathrm{whole},p}$ denote the density before any vertices are removed. 
We first compute the relative density $r_{a,p}=\rho_{a,p}/\rho_{\mathrm{whole},p}$. For a group $G$ consisting of one dataset and one value of $k$, let $\mathcal{P}_G$ be its set of
query meta-paths. We report their geometric mean as the normalized density gain $\mathrm{NGD}_{a,G}=\exp\bigl(\frac{1}{|\mathcal{P}_G|}
\sum_{p\in\mathcal{P}_G}\log r_{a,p}\bigr)$.

\begin{table}[t]
    \centering
    \caption{Normalized Density Gain of the Compared Methods}
    \label{tab:density-gain}
    \vspace{-2ex}
    \footnotesize
    \setlength{\tabcolsep}{4.0pt}
    \begin{tabular}{c|c|c|c|c|c|c|c|c}
        \hline
        \multirow{2}{*}{Method} & \multicolumn{2}{c|}{MovieLens} & \multicolumn{2}{c|}{DBLP} & \multicolumn{2}{c|}{DBpedia} & \multicolumn{2}{c}{Freebase} \\
        \cline{2-9}
        & $k=3$ & $k=4$ & $k=3$ & $k=4$ & $k=3$ & $k=4$ & $k=3$ & $k=4$ \\
        \hline
        \MAvgP & 1.27 & 1.05 & 0.77 & 0.37 & 2.12 & 2.00 & 7.90 & 2.23 \\
        \hline
        \iBF & 1.03 & 1.00 & 1.13 & 1.00 & 2.71 & 1.61 & 4.98 & 1.37 \\
        \hline
        \advexact & \textbf{1.46} & \textbf{1.31} & 1.94 & 1.37 & 5.80 & 5.08 & 32.90 & 13.93 \\
        \hline
        \ours & \textbf{1.46} & \textbf{1.31} & \textbf{1.96} & \textbf{1.38} & \textbf{5.82} & \textbf{5.15} & \textbf{33.21} & \textbf{14.01} \\
        \hline
    \end{tabular}
\end{table}

Table~\ref{tab:density-gain} shows that \ours matches \advexact on MovieLens after rounding and yields slightly larger density gains in all six datasets. 
For example, on DBpedia, the gains increase from 5.80 to 5.82 for $k=3$ and from 5.08 to 5.15
for $k=4$. 
Both methods optimize the same DPpS density and should therefore
return the same best density. The small difference in Table
\ref{tab:density-gain} comes from floating-point threshold updates in the implementation of \advexact. The resulting
precision loss can stop its fixed-$\mathbf{M}$ iteration before the best
density is reached. We verify the solutions returned by \ours using
exact pruning and fixed-$\mathbf{M}$ certificates, indicating that
\ours recovers solutions missed by the reference implementation.

Compared with \MAvgP and \iBF, \ours achieves higher NGD because it directly optimizes the DPpS density. \MAvgP maximizes the average
number of meta-path instances without normalizing by the geometric
mean of the selected vertex-set sizes, whereas \iBF focuses on
butterfly-core cohesion and local support. Their returned subgraphs
therefore need not maximize the density of complete meta-path
instances over all positions of the query meta-path. This difference
is more evident on the larger and more heterogeneous DBpedia and
Freebase graphs.

\begin{figure}[t]
    \centering
    \begin{minipage}[t]{0.49\columnwidth}
        \centering
        \includegraphics[width=\linewidth]{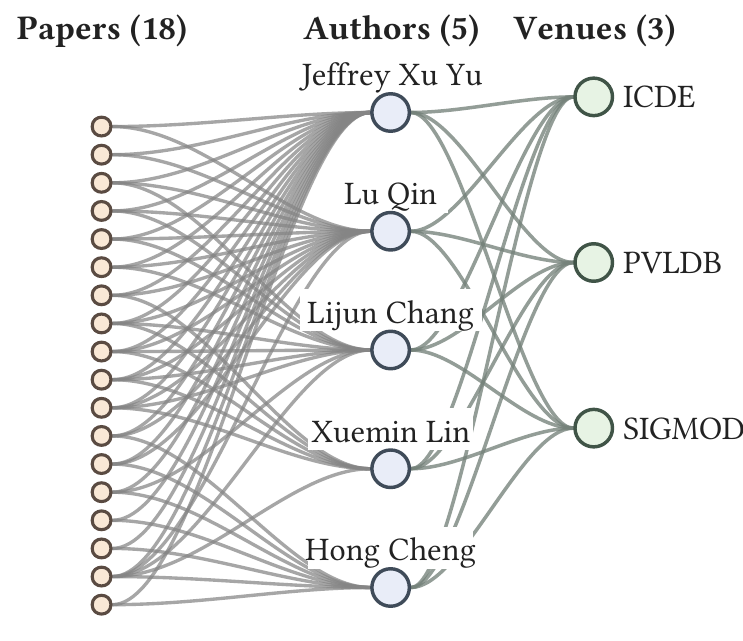}\\[-0.8ex]
        {\footnotesize\textbf{(a)} iBF, $\rho=25.529$}
    \end{minipage}\hfill
    \begin{minipage}[t]{0.49\columnwidth}
        \centering
        \includegraphics[width=\linewidth]{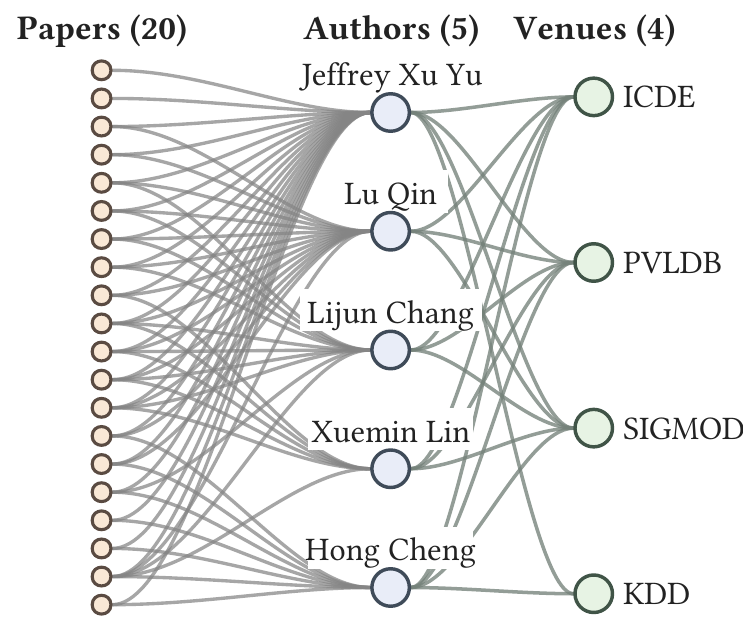}\\[-0.8ex]
        {\footnotesize\textbf{(b) \ours}, $\rho=26.873$}
    \end{minipage}
    \Description{Two side-by-side Paper--Author--Venue networks use identical
    node slots. The iBF result contains eighteen papers, five labeled authors,
    and three labeled venues. The \ours result adds two papers and KDD.}
    \caption{Case study on a DBLP Paper--Author--Venue query}
    \label{fig:case-study-comparison}
\end{figure}

\stitle{Exp-3: Effectiveness of Qualitative Case Study.}
As a qualitative case study, Fig.~\ref{fig:case-study-comparison} visualizes \ours and \iBF on an ego-centric  Paper--Author--Venue subgraph around author `Jeffrey Xu Yu', extracted from 2010--2024 DBLP. The subgraphs are searched from DBLP records from 2010 to 2024 for the four venues: SIGMOD, VLDB, ICDE, and KDD.
We report the \iBF result with the highest DPpS density over all butterfly thresholds. Both methods deliver the same five-author core. However, \ours additionally retains the venue KDD and two papers, adding 33 meta-path instances and increasing the density from 25.529 to 26.873. This case study illustrates that vertices with limited local support may still jointly improve the
global multipartite density.

\stitle{Exp-4: Cybersecurity User Grouping.}
We further evaluate application-level group quality on Cisco g21, which contains non-overlapping ground-truth user groups in a $\mathsf{User}$--$\mathsf{Port}$--$\mathsf{Protocol}$--$\mathsf{Server}$ HIN. We repeatedly extract the users in the densest $P$-partite
subgraph as a discovered group and report the average F1 score against the ground-truth groups. We vary the packet threshold $\tau$ and retain communication relations involving at least $\tau$ packets.

Table~\ref{tab:cisco-f1} shows that \ours matches \advexact at five
of the six packet thresholds and improves the F1 from $0.505$ to
$0.633$ at $\tau=5{,}000$. These results indicate that the
efficiency-oriented transformations of \ours do not sacrifice
application-level quality.

\stitle{Exp-5: Ablation Studies.}
To isolate the effect of each proposed component, we conduct the ablation study on a controlled variant of \ours that uses the same network-flow algorithm as \advexact for all fixed-$\mathbf{M}$ subproblems. 
The configuration with all evaluated components enabled is denoted by \Full.
Each ablated variant disables one component from \Full while keeping all other components unchanged. Specifically, \NoPrim
disables primitive count-vector canonicalization; \NoBoxUB disables all box-level upper bounds; \NoWS disables both warm-up and exact seed scheduling; and \NoGroup disables projection grouping in fixed-$\mathbf{M}$ solving.
Table~\ref{tab:ablation-runtime} reports the average running time over the meta-path queries for each dataset and each value $k \in \{3, 4\}$. 
To summarize the overall effect of a component, we additionally compare the accumulated running time of its variant with that of \Full.

\begin{table}[t]
    \centering
    \caption{F1 Scores on G21 under Different Packet Thresholds}
    \label{tab:cisco-f1}
    \vspace{-2ex}
    \footnotesize
    \setlength{\tabcolsep}{4.0pt}
    \renewcommand{\arraystretch}{1.14}
    \begin{tabular}{c|c|c|c|c|c|c}
        \hline
        Method & $1$ & $10$ & $100$ & $1{,}000$ & $5{,}000$ & $10{,}000$ \\
        \hline
        \MAvgP & 0.435 & 0.435 & 0.435 & 0.375 & 0.241 & 0.248 \\
        \hline
        \iBF & 0.435 & 0.435 & 0.435 & 0.402 & 0.402 & 0.402 \\
        \hline
        \advexact & \textbf{0.615} & \textbf{0.615} & \textbf{0.615} & \textbf{0.519} & 0.505 & \textbf{0.514} \\
        \hline
        \ours & \textbf{0.615} & \textbf{0.615} & \textbf{0.615} & \textbf{0.519} & \textbf{0.633} & \textbf{0.514} \\
        \hline
    \end{tabular}
\end{table}

\begin{table}[t]
    \centering
    \caption{Arithmetic Mean of Per-query Running Times (s)}
    \label{tab:ablation-runtime}
    \vspace{-2ex}
    \footnotesize
    \setlength{\tabcolsep}{4.0pt}
    \resizebox{\linewidth}{!}{
    \begin{tabular}{c| r| r|r|r|r|r|r|r}
        \hline
        \multirow{2}{*}{Method} & \multicolumn{2}{c|}{MovieLens} & \multicolumn{2}{c|}{DBLP} & \multicolumn{2}{c|}{DBpedia} & \multicolumn{2}{c}{Freebase} \\
        \cline{2-9}
        & $k=3$ & $k=4$ & $k=3$ & $k=4$ & $k=3$ & $k=4$ & $k=3$ & $k=4$ \\
        \hline
        \Full & \textbf{11.70} & \textbf{15.36} & \textbf{23.31} & \textbf{19.54} & \textbf{22.48} & \textbf{68.32} & \textbf{812.20} & \textbf{188.77} \\
        \hline
        \NoPrim & 12.08 & 17.23 & 55.50 & 31.61 & 27.84 & 93.89 & 923.44 & 235.87 \\
        \hline
        \NoBoxUB & 11.73 & 16.10 & 30.15 & 24.48 & 26.10 & 104.66 & 1152.36 & 299.52 \\
        \hline
        \NoWS & 12.07 & 17.97 & 23.62 & 441.80 & 24.54 & 221.29 & 1077.63 & 3045.11 \\
        \hline
        \NoGroup & 20.15 & 96.32 & 101.42 & 51.15 & 46.21 & 79.05 & 1408.02 & 299.78 \\
        \hline
    \end{tabular}}
\end{table}

The three search-side techniques prune the search space at different stages.
Warm-up and exact seed scheduling make useful lower bounds and strict certificates available early in the search. 
Their effect is limited on most $k=3$ queries but becomes remarkable on the longer $k=4$ queries: \NoWS is the slowest variant on
DBLP, DBpedia, and Freebase for $k=4$. 
Aggregated over all evaluated queries, disabling the two techniques increases the running time to
$3.76\times$ that of \Full. 
Box upper bounds subsequently use the current lower bound to discard count-vector regions without examining their members individually, whereas primitive canonicalization avoids reprocessing count vectors that represent the same iRM-set. The more moderate slowdowns of \NoBoxUB and \NoPrim in Table~\ref{tab:ablation-runtime} reflect their complementary roles: region-level pruning removes unpromising portions of the search space, while canonicalization removes duplicate representations.

A separate source of cost arises after a count vector survives the search-side pruning and requires an exact fixed-$\mathbf{M}$ solve. We observe the benefit of projection grouping across all four datasets
for $k=3$, which remains substantial on several $k=4$ queries.
Disabling projection grouping increases the accumulated running time to $1.70\times$ that of \Full.
This result confirms that reducing the network-flow instance for each remaining exact solve is important, even after the number of examined count vectors has been reduced. 
Although some individual ablations cause only
moderate slowdowns, we find that removing all evaluated components together causes most query workloads to exceed the time limit. This result indicates that these components are complementary and have a substantially stronger combined effect. 
Overall, the ablated results show that our proposed components both avoid unnecessary fixed-$\mathbf{M}$ subproblems and reduce the cost of fixed-$\mathbf{M}$ subproblems that remain.

\begin{figure}[t]
    \centering
    \includegraphics[width=0.95\columnwidth]{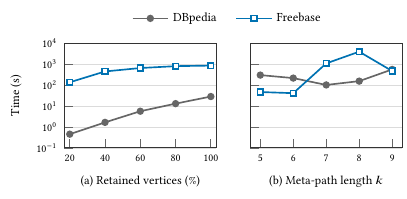}
    \Description{Two logarithmic line charts show the mean running times
    of BoxDPpS on DBpedia and Freebase as the retained vertex percentage
    increases from 20 to 100 and as the meta-path length increases from
    five to nine.}
    \vspace{-3ex}
    \caption{Scalability of \ours w.r.t. data size and
    $k$}
    \label{fig:scalability}
\end{figure}

\stitle{Exp-6: Scalability w.r.t. data size.}
We study the scalability of \ours, by varying the graph size.
We independently sample percentages of $\{20\%, 40\%, 60\%, 80\%, 100\%\}$ vertices from each vertex type and construct the corresponding induced HIN. 
For each percentage below $100\%$, we repeat the sampling using ten fixed random seeds; the $100\%$ setting uses the complete dataset. We evaluate the $k = 3$ query set at every graph scale. 
Each point in Fig.~\ref{fig:scalability}(a) reports the average per-query running time.

Fig.~\ref{fig:scalability}(a) shows that the running time of
\ours increases steadily with the retained data size. Freebase
requires more time because its sampled query graphs contain
substantially more vertices and edges than those of DBpedia.
Nevertheless, \ours completes all queries at every tested scale,
including the complete datasets. This result demonstrates that
\ours scales to the full DBpedia and Freebase workloads considered
in this experiment.

\stitle{Exp-7: Scalability w.r.t. meta-path length.}
We further evaluate \ours on DBpedia and Freebase using meta-paths
with $k\in\{5,6,7,8,9\}$. For each dataset and each value of $k$, we
select structurally diverse queries using fixed criteria based on the
induced graph size, the number of meta-path instances, and the
estimated count-vector search space. Each point in
Fig.~\ref{fig:scalability}(b) reports the average per-query running time.

The running time is non-monotonic in $k$. Recall that for DPpS, longer meta-paths tend to induce smaller connected $P$-partite graphs
because distant vertex types are more weakly related~\cite{ChenLZLXL23};
the decreasing vertex and edge counts in our workloads exhibit the
same size effect. For \ours, however, each additional type also raises
the dimension of the count-vector search. The frontier may consequently
visit more boxes, and testing a lifted exact certificate on one box may inspect up to $2^k$ log-box corners. The reduced graph size lowers instance-processing and fixed-$\mathbf{M}$ solving costs, whereas the higher-dimensional box search increases frontier and pruning costs.
The interaction between these opposing factors explains the non-monotonic curves and shows that the induced query structure, rather than $k$ alone, determines the practical workload. Overall, \ours processes all tested meta-paths with up to nine vertex-type positions.

\section{Related Work}
\label{sec:related-work}

\stitle{Densest Subgraph Search.} 
Densest subgraph search is a fundamental graph mining problem that seeks a vertex-induced subgraph with maximum density, which has attracted extensive attention from the community~\cite{JiangYLWHC25,Teng0ZLFW24,ZhangXWYJ22,OettershagenWG24,SahaK0L23a,GalimbertiBG17,WangYL24,LuoLGX25,JiangHZWFLC25,LiuGW22,KimKHKJK26,SunHLCX22,ChenLXWL25}.
The classical edge-density formulation admits exact max-flow-based solutions~\cite{Goldberg:CSD-84-171} and greedy peeling approximations~\cite{Charikar00}. 
Subsequent studies have improved scalability and broadened the objective family. For example, core-based reduction and localization techniques accelerate exact and approximate densest subgraph discovery on large graphs~\cite{FangYCLL19}, while supermodularity, peeling, and flow have been studied in a unified framework~\cite{DBLP:conf/soda/ChekuriQT22}. 
Directed densest subgraph search is closely related to our setting because it optimizes density over two vertex sets instead of one. Recent directed methods improve efficiency using core-style pruning, exact/approximate algorithms, and convex-programming formulations~\cite{MaFCL0020, MaFCLH22}. Beyond edge density, higher-order density objectives use motifs or cliques to reveal stronger cohesion. The $k$-clique densest subgraph problem generalizes edge density to clique counts~\cite{DBLP:conf/www/Tsourakakis15a}, and recent counting-based approaches reduce expensive clique enumeration when optimizing such objectives~\cite{ZhouGF024}.
However, these studies are designed mainly for homogeneous, directed, or higher-order homogeneous graphs.
The most relevant work is the densest $P$-partite subgraph search problem in HINs~\cite{ChenLZLXL23}, which provides exact and approximate algorithms based on iRM-set enumeration and fixed-$\mathbf{M}$ min-cut subproblems. Our work keeps the same DPpS objective and exactness guarantee, but provides an efficient solution by pruning count space and reducing the cost of surviving fixed-$\mathbf{M}$ solves.

\noindent\textbf{Cohesive Subgraph Search in HINs.} 
HINs model multi-typed objects and relations, where meta-paths are widely used to express typed semantics.
Early HIN studies~\cite{SunHYYW11, ShiKHYW14} define meta-path-based similarity and relevance measures as the semantic foundations of cohesive subgraphs over HINs. 
Cohesive subgraph and community search in HINs adapt classic notions such as core, truss, clique, and relational constraints to typed networks. Meta-path-based community search identifies cohesive communities containing query vertices~\cite{FangYZLC20},  and star-schema HINs further support hub-centered community models~\cite{DBLP:journals/pvldb/JiangFMCL22}.
Truss-based HIN models strengthen cohesion using triangle-like typed structures~\cite{YangF00F20}, while relational community search handles dynamic HINs with relation constraints~\cite{JianWC20}. Motif-clique models capture richer typed patterns beyond simple paths~\cite{DBLP:conf/icde/HuCCSFL19}. 
Recent extensions incorporate additional semantics, including influence~\cite{DBLP:journals/pvldb/ZhouFLY23}, structural similarity and vertex roles~\cite{WangFL25}, and temporal proximity~\cite{DBLP:journals/pvldb/TangLCZL25}. 
These models are effective for query-driven or constraint-based HIN mining, but they usually require query vertices, thresholds, influence or attribute constraints, temporal windows, or predefined structural roles. They also often focus on one target type or on satisfying a cohesiveness constraint. In contrast, DPpS is a global density maximization problem over all vertex types along a meta-path.

\section{Conclusion}
\label{sec:conclusion}

We study exact densest $P$-partite subgraph mining in HINs under the DPpS density objective. The main bottleneck is the combination of high-dimensional count-vector search and repeated fixed-$\mathbf{M}$ solves over many meta-path instances. We propose a box-level exact search that prunes count regions by safe upper bounds and lifted exact pruning conditions, deduplicates repeated iRM-sets by primitive count-vector keys, and uses bounded warm-up and seed scheduling to improve $\rho_{\mathrm{best}}$ early. We further introduce a multiplicity-aware exact solution via terminal-twin compression and projection grouping, thereby reducing the flow network while preserving the original auxiliary objective. Together, these techniques provide an exact solution that reduces redundant search without weakening the optimality guarantee. For the query meta-paths completed, \ours achieves an arithmetic mean speedup of $27.04\times$ over the state-of-the-art approach \advexact.

\bibliographystyle{ACM-Reference-Format}
\balance
\bibliography{sample}

@article{ChenLZLXL23,
  author       = {Lu Chen and
                  Chengfei Liu and
                  Rui Zhou and
                  Kewen Liao and
                  Jiajie Xu and
                  Jianxin Li},
  title        = {Densest Multipartite Subgraph Search in Heterogeneous Information
                  Networks},
  journal      = {Proc. {VLDB} Endow.},
  volume       = {17},
  number       = {4},
  pages        = {699--711},
  year         = {2023}
}

@inproceedings{Charikar00,
  author       = {Moses Charikar},
  editor       = {Klaus Jansen and
                  Samir Khuller},
  title        = {Greedy approximation algorithms for finding dense components in a
                  graph},
  booktitle    = {Approximation Algorithms for Combinatorial Optimization, Third International
                  Workshop, {APPROX} 2000, Saarbr{\"{u}}cken, Germany, September
                  5-8, 2000, Proceedings},
  series       = {Lecture Notes in Computer Science},
  volume       = {1913},
  pages        = {84--95},
  publisher    = {Springer},
  year         = {2000},
}

@article{FangYCLL19,
  author       = {Yixiang Fang and
                  Kaiqiang Yu and
                  Reynold Cheng and
                  Laks V. S. Lakshmanan and
                  Xuemin Lin},
  title        = {Efficient Algorithms for Densest Subgraph Discovery},
  journal      = {Proc. {VLDB} Endow.},
  volume       = {12},
  number       = {11},
  pages        = {1719--1732},
  year         = {2019},
}

@article{SunHYYW11,
  author       = {Yizhou Sun and
                  Jiawei Han and
                  Xifeng Yan and
                  Philip S. Yu and
                  Tianyi Wu},
  title        = {PathSim: Meta Path-Based Top-K Similarity Search in Heterogeneous
                  Information Networks},
  journal      = {Proc. {VLDB} Endow.},
  volume       = {4},
  number       = {11},
  pages        = {992--1003},
  year         = {2011},
}

@article{ZhouGF024,
  author       = {Yingli Zhou and
                  Qingshuo Guo and
                  Yixiang Fang and
                  Chenhao Ma},
  title        = {A Counting-based Approach for Efficient k-Clique Densest Subgraph
                  Discovery},
  journal      = {Proc. {ACM} Manag. Data},
  volume       = {2},
  number       = {3},
  pages        = {119},
  year         = {2024},
}

@article{ShiKHYW14,
  author       = {Chuan Shi and
                  Xiangnan Kong and
                  Yue Huang and
                  Philip S. Yu and
                  Bin Wu},
  title        = {HeteSim: {A} General Framework for Relevance Measure in Heterogeneous
                  Networks},
  journal      = {{IEEE} Trans. Knowl. Data Eng.},
  volume       = {26},
  number       = {10},
  pages        = {2479--2492},
  year         = {2014},
}

@article{FangYZLC20,
  author       = {Yixiang Fang and
                  Yixing Yang and
                  Wenjie Zhang and
                  Xuemin Lin and
                  Xin Cao},
  title        = {Effective and Efficient Community Search over Large Heterogeneous
                  Information Networks},
  journal      = {Proc. {VLDB} Endow.},
  volume       = {13},
  number       = {6},
  pages        = {854--867},
  year         = {2020},
}

@inproceedings{YangF00F20,
  author       = {Yixing Yang and
                  Yixiang Fang and
                  Xuemin Lin and
                  Wenjie Zhang},
  title        = {Effective and Efficient Truss Computation over Large Heterogeneous
                  Information Networks},
  booktitle    = {36th {IEEE} International Conference on Data Engineering, {ICDE} 2020,
                  Dallas, TX, USA, April 20-24, 2020},
  pages        = {901--912},
  publisher    = {{IEEE}},
  year         = {2020},
}

@article{JianWC20,
  author       = {Xun Jian and
                  Yue Wang and
                  Lei Chen},
  title        = {Effective and Efficient Relational Community Detection and Search
                  in Large Dynamic Heterogeneous Information Networks},
  journal      = {Proc. {VLDB} Endow.},
  volume       = {13},
  number       = {10},
  pages        = {1723--1736},
  year         = {2020},
}

@article{WangFL25,
  author       = {Shu Wang and
                  Yixiang Fang and
                  Wensheng Luo},
  title        = {Searching and Detecting Structurally Similar Communities in Large
                  Heterogeneous Information Networks},
  journal      = {Proc. {VLDB} Endow.},
  volume       = {18},
  number       = {5},
  pages        = {1425--1438},
  year         = {2025},
}

@inproceedings{MaFCL0020,
  author       = {Chenhao Ma and
                  Yixiang Fang and
                  Reynold Cheng and
                  Laks V. S. Lakshmanan and
                  Wenjie Zhang and
                  Xuemin Lin},
  editor       = {David Maier and
                  Rachel Pottinger and
                  AnHai Doan and
                  Wang{-}Chiew Tan and
                  Abdussalam Alawini and
                  Hung Q. Ngo},
  title        = {Efficient Algorithms for Densest Subgraph Discovery on Large Directed
                  Graphs},
  booktitle    = {Proceedings of the 2020 International Conference on Management of
                  Data, {SIGMOD} Conference 2020, online conference [Portland, OR, USA],
                  June 14-19, 2020},
  pages        = {1051--1066},
  publisher    = {{ACM}},
  year         = {2020},
}

@inproceedings{MaFCLH22,
  author       = {Chenhao Ma and
                  Yixiang Fang and
                  Reynold Cheng and
                  Laks V. S. Lakshmanan and
                  Xiaolin Han},
  editor       = {Zachary G. Ives and
                  Angela Bonifati and
                  Amr El Abbadi},
  title        = {A Convex-Programming Approach for Efficient Directed Densest Subgraph
                  Discovery},
  booktitle    = {{SIGMOD} '22: International Conference on Management of Data, Philadelphia,
                  PA, USA, June 12 - 17, 2022},
  pages        = {845--859},
  publisher    = {{ACM}},
  year         = {2022},
}

@article{Hochbaum08,
  author       = {Dorit S. Hochbaum},
  title        = {The Pseudoflow Algorithm: {A} New Algorithm for the Maximum-Flow Problem},
  journal      = {Oper. Res.},
  volume       = {56},
  number       = {4},
  pages        = {992--1009},
  year         = {2008},
}

@inproceedings{MadaniAG22,
  author       = {Omid Madani and
                  Sai Ankith Averineni and
                  Shashidhar Gandham},
  editor       = {Andrew H. Sung and
                  Rakesh M. Verma and
                  Roland H. C. Yap},
  title        = {A Dataset of Networks of Computing Hosts},
  booktitle    = {IWSPA@CODASPY 2022: Proceedings of the 2022 {ACM} on International
                  Workshop on Security and Privacy Analytics, Baltimore, MD, USA, April
                  27, 2022},
  pages        = {100--104},
  publisher    = {{ACM}},
  year         = {2022}
}

@article{DongHYZX21,
  author       = {Zheng Dong and
                  Xin Huang and
                  Guorui Yuan and
                  Hengshu Zhu and
                  Hui Xiong},
  title        = {Butterfly-Core Community Search over Labeled Graphs},
  journal      = {Proc. {VLDB} Endow.},
  volume       = {14},
  number       = {11},
  pages        = {2006--2018},
  year         = {2021}
}

@inproceedings{DBLP:conf/soda/ChekuriQT22,
  author       = {Chandra Chekuri and
                  Kent Quanrud and
                  Manuel R. Torres},
  editor       = {Joseph (Seffi) Naor and
                  Niv Buchbinder},
  title        = {Densest Subgraph: Supermodularity, Iterative Peeling, and Flow},
  booktitle    = {Proceedings of the 2022 {ACM-SIAM} Symposium on Discrete Algorithms,
                  {SODA} 2022, Virtual Conference / Alexandria, VA, USA, January 9 -
                  12, 2022},
  pages        = {1531--1555},
  publisher    = {{SIAM}},
  year         = {2022}
}

@techreport{Goldberg:CSD-84-171,
    Author = {Goldberg, A. V.},
    Title = {Finding a Maximum Density Subgraph},
    Institution = {EECS Department, University of California, Berkeley},
    Year = {1984},
    Number = {UCB/CSD-84-171},
    Url = {https://www2.eecs.berkeley.edu/Pubs/TechRpts/1984/5956.html}
}

@inproceedings{DBLP:conf/www/Tsourakakis15a,
  author       = {Charalampos E. Tsourakakis},
  editor       = {Aldo Gangemi and
                  Stefano Leonardi and
                  Alessandro Panconesi},
  title        = {The K-clique Densest Subgraph Problem},
  booktitle    = {Proceedings of the 24th International Conference on World Wide Web,
                  {WWW} 2015, Florence, Italy, May 18-22, 2015},
  pages        = {1122--1132},
  publisher    = {{ACM}},
  year         = {2015}
}

@article{DBLP:journals/pvldb/JiangFMCL22,
  author       = {Yangqin Jiang and
                  Yixiang Fang and
                  Chenhao Ma and
                  Xin Cao and
                  Chunshan Li},
  title        = {Effective Community Search over Large Star-Schema Heterogeneous Information
                  Networks},
  journal      = {Proc. {VLDB} Endow.},
  volume       = {15},
  number       = {11},
  pages        = {2307--2320},
  year         = {2022}
}

@inproceedings{DBLP:conf/icde/HuCCSFL19,
  author       = {Jiafeng Hu and
                  Reynold Cheng and
                  Kevin Chen{-}Chuan Chang and
                  Aravind Sankar and
                  Yixiang Fang and
                  Brian Y. H. Lam},
  title        = {Discovering Maximal Motif Cliques in Large Heterogeneous Information
                  Networks},
  booktitle    = {35th {IEEE} International Conference on Data Engineering, {ICDE} 2019,
                  Macao, China, April 8-11, 2019},
  pages        = {746--757},
  publisher    = {{IEEE}},
  year         = {2019}
}

@article{DBLP:journals/pvldb/ZhouFLY23,
  author       = {Yingli Zhou and
                  Yixiang Fang and
                  Wensheng Luo and
                  Yunming Ye},
  title        = {Influential Community Search over Large Heterogeneous Information
                  Networks},
  journal      = {Proc. {VLDB} Endow.},
  volume       = {16},
  number       = {8},
  pages        = {2047--2060},
  year         = {2023}
}

@article{DBLP:journals/pvldb/TangLCZL25,
  author       = {Yifu Tang and
                  Chengfei Liu and
                  Lu Chen and
                  Rui Zhou and
                  Jianxin Li},
  title        = {Finding Time-Proximity Communities in Temporal Heterogeneous Information
                  Networks},
  journal      = {Proc. {VLDB} Endow.},
  volume       = {18},
  number       = {13},
  pages        = {5740--5752},
  year         = {2025}
}

@misc{HIN-Datasets,
url={https://github.com/librahu/HIN-Datasets-for-Recommendation-and-Network-Embedding}, 
journal={GitHub - Librahu/hin-datasets-for-recommendation-and-network-embedding: Heterogeneous Information Network datasets for recommendation and network embedding}}

@article{JiangYLWHC25,
  author       = {Jiaxin Jiang and
                  Siyuan Yao and
                  Yuchen Li and
                  Qiange Wang and
                  Bingsheng He and
                  Min Chen},
  title        = {Dupin: {A} Parallel Framework for Densest Subgraph Discovery in Fraud
                  Detection on Massive Graphs},
  journal      = {Proc. {ACM} Manag. Data},
  volume       = {3},
  number       = {3},
  pages        = {150:1--150:26},
  year         = {2025}
}

@inproceedings{Teng0ZLFW24,
  author       = {Siyi Teng and
                  Jiadong Xie and
                  Fan Zhang and
                  Can Lu and
                  Juntao Fang and
                  Kai Wang},
  title        = {Optimizing Network Resilience via Vertex Anchoring},
  booktitle    = {Proceedings of the {ACM} on Web Conference 2024, {WWW} 2024, Singapore,
                  May 13-17, 2024},
  pages        = {606--617},
  publisher    = {{ACM}},
  year         = {2024}
}

@article{JiangYCHNLSL25,
  author       = {Jiaxin Jiang and
                  Siyuan Yao and
                  Yuhang Chen and
                  Bingsheng He and
                  Yudong Niu and
                  Yuchen Li and
                  Shixuan Sun and
                  Yongchao Liu},
  title        = {Community Detection in Heterogeneous Information Networks Without
                  Materialization},
  journal      = {Proc. {ACM} Manag. Data},
  volume       = {3},
  number       = {3},
  pages        = {139:1--139:27},
  year         = {2025}
}

@article{ZhangXWYJ22,
  author       = {Fan Zhang and
                  Jiadong Xie and
                  Kai Wang and
                  Shiyu Yang and
                  Yu Jiang},
  title        = {Discovering key users for defending network structural stability},
  journal      = {World Wide Web},
  volume       = {25},
  number       = {2},
  pages        = {679--701},
  year         = {2022}
}

@inproceedings{OettershagenWG24,
  author       = {Lutz Oettershagen and
                  Honglian Wang and
                  Aristides Gionis},
  title        = {Finding Densest Subgraphs with Edge-Color Constraints},
  booktitle    = {Proceedings of the {ACM} on Web Conference 2024, {WWW} 2024, Singapore,
                  May 13-17, 2024},
  pages        = {936--947},
  publisher    = {{ACM}},
  year         = {2024}
}

@inproceedings{SahaK0L23a,
  author       = {Arkaprava Saha and
                  Xiangyu Ke and
                  Arijit Khan and
                  Cheng Long},
  title        = {Most Probable Densest Subgraphs},
  booktitle    = {39th {IEEE} International Conference on Data Engineering, {ICDE} 2023,
                  Anaheim, CA, USA, April 3-7, 2023},
  pages        = {1447--1460},
  publisher    = {{IEEE}},
  year         = {2023}
}

@inproceedings{GalimbertiBG17,
  author       = {Edoardo Galimberti and
                  Francesco Bonchi and
                  Francesco Gullo},
  title        = {Core Decomposition and Densest Subgraph in Multilayer Networks},
  booktitle    = {Proceedings of the 2017 {ACM} on Conference on Information and Knowledge
                  Management, {CIKM} 2017, Singapore, November 06 - 10, 2017},
  pages        = {1807--1816},
  publisher    = {{ACM}},
  year         = {2017}
}

@article{WangYL24,
  author       = {Kaixin Wang and
                  Kaiqiang Yu and
                  Cheng Long},
  title        = {Efficient k-Clique Listing: An Edge-Oriented Branching Strategy},
  journal      = {Proc. {ACM} Manag. Data},
  volume       = {2},
  number       = {1},
  pages        = {7:1--7:26},
  year         = {2024}
}

@article{LuoLGX25,
  author       = {Chengyang Luo and
                  Qing Liu and
                  Yunjun Gao and
                  Jianliang Xu},
  title        = {Synergetic Community Search over Large Multilayer Graphs},
  journal      = {Proc. {VLDB} Endow.},
  volume       = {18},
  number       = {5},
  pages        = {1412--1424},
  year         = {2025}
}

@article{JiangHZWFLC25,
  author       = {Jiawei Jiang and
                  Hao Huang and
                  Zhigao Zheng and
                  Yi Wei and
                  Fangcheng Fu and
                  Xiaosen Li and
                  Bin Cui},
  title        = {Detecting and Analyzing Motifs in Large-Scale Online Transaction Networks},
  journal      = {{IEEE} Trans. Knowl. Data Eng.},
  volume       = {37},
  number       = {2},
  pages        = {584--596},
  year         = {2025}
}

@inproceedings{KimKHKJK26,
  author       = {Song Kim and
                  Dahee Kim and
                  Taejoon Han and
                  Junghoon Kim and
                  Hyun Ji Jeong and
                  Jungeun Kim},
  title        = {Efficient Locality-based Indexing for Cohesive Subgraphs Discovery
                  in Hypergraphs},
  booktitle    = {Proceedings 29th International Conference on Extending Database Technology,
                  {EDBT} 2026, Tampere, Finland, March 24-27, 2026},
  pages        = {170--182},
  publisher    = {OpenProceedings.org},
  year         = {2026}
}

@article{LiuGW22,
  author       = {Xuanming Liu and
                  Tingjian Ge and
                  Yinghui Wu},
  title        = {A Stochastic Approach to Finding Densest Temporal Subgraphs in Dynamic
                  Graphs},
  journal      = {{IEEE} Trans. Knowl. Data Eng.},
  volume       = {34},
  number       = {7},
  pages        = {3082--3094},
  year         = {2022}
}

@article{SunHLCX22,
  author       = {Longxu Sun and
                  Xin Huang and
                  Rong{-}Hua Li and
                  Byron Choi and
                  Jianliang Xu},
  title        = {Index-Based Intimate-Core Community Search in Large Weighted Graphs},
  journal      = {{IEEE} Trans. Knowl. Data Eng.},
  volume       = {34},
  number       = {9},
  pages        = {4313--4327},
  year         = {2022}
}

@inproceedings{HuangZCSML16,
  author       = {Zhipeng Huang and
                  Yudian Zheng and
                  Reynold Cheng and
                  Yizhou Sun and
                  Nikos Mamoulis and
                  Xiang Li},
  title        = {Meta Structure: Computing Relevance in Large Heterogeneous Information
                  Networks},
  booktitle    = {Proceedings of the 22nd {ACM} {SIGKDD} International Conference on
                  Knowledge Discovery and Data Mining, San Francisco, CA, USA, August
                  13-17, 2016},
  pages        = {1595--1604},
  publisher    = {{ACM}},
  year         = {2016}
}

@inproceedings{ChenLXWL25,
  author       = {Xin Chen and
                  Wenqing Lin and
                  Haoxuan Xie and
                  Sibo Wang and
                  Siqiang Luo},
  title        = {Finding Near-Optimal Maximum Set of Disjoint {\textdollar}k{\textdollar}-Cliques
                  in Real-World Social Networks},
  booktitle    = {41st {IEEE} International Conference on Data Engineering, {ICDE} 2025,
                  Hong Kong, May 19-23, 2025},
  pages        = {3316--3328},
  publisher    = {{IEEE}},
  year         = {2025}
}

\end{document}